%% file: Paper.tex
\newcommand*{\ATLASLATEXPATH}{latex/}
\documentclass[cernpreprint,txfonts,UKenglish,texlive=2011]{\ATLASLATEXPATH atlasdoc}
\pdfoutput=1
\usepackage{\ATLASLATEXPATH atlaspackage}
\usepackage{\ATLASLATEXPATH atlasbiblatex}

\usepackage{\ATLASLATEXPATH atlascontribute}

\usepackage{\ATLASLATEXPATH atlasphysics}

\usepackage{multirow}

\graphicspath{{logos/}{figures/}}

\usepackage{Paper-defs}

\hypersetup{pdftitle={ATLAS document},pdfauthor={The ATLAS Collaboration}}
\input{Paper-metadata}

\begin{document}

\maketitle

\tableofcontents


\clearpage



\input introduction

\input theory

\input method

\input atlas

\input samples

\input analysis

\input fit

\input results

\input conclusions

\input Acknowledgements

\printbibliography

\newpage 
\input{atlas_authlist}


%

\end{document}

%% file: Paper-metadata.tex
\AtlasTitle{Test of CP Invariance in  vector-boson fusion production of the Higgs boson 
using the Optimal Observable method in the ditau decay channel with the ATLAS detector}
\author{The ATLAS Collaboration}

\AtlasRefCode{HIGG-2015-06}

\AtlasNote{ATL-COM-PHYS-2015-1139}

\PreprintIdNumber{CERN-EP-2016-002}

\AtlasDate{\today}

 \AtlasJournal{EPJC}
 \AtlasJournalRef{Eur. Phys. J. C76 (2016) 658}
 \AtlasDOI{10.1140/epjc/s10052-016-4499-5}

\AtlasAbstract{%
A test of CP invariance in Higgs boson production via vector-boson fusion using 
the method of the \textit{Optimal Observable} is presented. The analysis exploits the decay 
mode of the Higgs boson into a pair of $\tau$ leptons and is based on 20.3 \ifb\ of proton--proton collision data at $\sqrt{s}$ = 8~\TeV\
collected  by the ATLAS experiment at the LHC. 
Contributions from CP-violating 
interactions between the Higgs boson and electroweak gauge bosons are described in an 
effective field theory framework, in which the strength of CP violation is governed by 
a single parameter \dtilde. The mean values and distributions of CP-odd observables
agree with the expectation in the Standard Model and show no sign of CP violation. 
The CP-mixing parameter $\dtilde$ is constrained to the interval $[-0.11,0.05]$ at 68\% 
confidence level, consistent with the Standard Model expectation of $\dtilde=0$.}

\AtlasCoverSupportingNote{Support Note: }{ https://cds.cern.ch/record/2002134}
\AtlasCoverCommentsDeadline{January 11th, 2015}

\AtlasCoverAnalysisTeam{
Attilio Andreazza, Quentin Buat, Elias Coniavitis, Vincent Croft, Noel Dawe, Sinead Farrington, Michel Janus, Florian Kiss,
Stanley Lai, Alessia Murrone, Elisabetta Pianori, Dirk Sammel, Christian Schillo, Ulrike Schnoor, Markus Schumacher, 
Martin Spangenberg, Daniele Zanzi
}

\AtlasCoverEdBoardMember{Dugan O'Neil~(chair), Evelin Meoni, Nikolaos Rompotis}

\AtlasCoverEgroupEditors{atlas-HIGG-2015-06-editors@cern.ch}

\AtlasCoverEgroupEdBoard{atlas-HIGG-2015-06-editorial-board@cern.ch}

%% file: introduction.tex
\section{Introduction}
\label{sec:intro}
The discovery of a Higgs boson by the ATLAS and CMS experiments~\cite{Aad:2012tfa,Chatrchyan:2012xdj} at the LHC~\cite{LHC}
offers a novel opportunity to search for new sources of CP violation in the 
interaction of the Higgs boson with other Standard Model (SM) particles.
C and CP violation is one of the three Sakharov conditions~\cite{Sakharov:1967dj,Sakharov:1979xt,Sakharov:1988pm}
needed to explain the observed baryon asymmetry of the universe. 
In the SM with massless neutrinos the only source of CP violation is the complex phase
in the quark mixing (CKM) matrix~\cite{Cabibbo:1963yz,Kobayashi:1973fv}. The measured 
size of the complex phase and the derived magnitude of CP violation in the early universe is insufficient 
to explain the observed value of the baryon asymmetry \cite{Ade:2015xua}
within the SM~\cite{Huet:1994jb,Gavela:1993ts}
and, most probably, new sources of CP violation beyond the SM need to be introduced.
No observable effect of CP violation is expected in the production or decay of the SM Higgs boson. Hence any observation of CP violation involving the observed Higgs boson  would be an unequivocal sign of physics beyond the SM.

The measured Higgs boson production cross sections, branching ratios and derived constraints on coupling-strength modifiers,
assuming the tensor structure of the SM, agree with the SM predictions~\cite{Khachatryan:2014jba,Aad:2015gba}.
Investigations of spin and CP quantum numbers in bosonic decay modes and measurements of anomalous couplings 
including CP-violating ones in the decay into a pair of massive electroweak gauge bosons 
show no hints of deviations from the tensor structure of the SM Higgs boson~\cite{Khachatryan:2014kca,Aad:2015mxa}.  
Differential cross-section measurements in the decay $H \rightarrow \gamma \gamma$ have been used to set 
limits on couplings including CP-violating ones in vector-boson fusion production in an effective field theory~\cite{Aad:2015tna}. 
However, the observables, including absolute event rates, used in that analysis were CP-even and hence not sensitive to the possible interference between the SM and CP-odd couplings and did not directly test CP invariance.
The observables used in this analysis are CP-odd and therefore sensitive to this interference and the measurement is designed as a direct test of CP invariance.

In this paper, a first direct test of CP invariance in Higgs boson production via  vector-boson fusion (VBF) is presented,
based on proton--proton collision data corresponding to an integrated luminosity of 20.3\,\fbs \,
collected with the ATLAS detector at $\sqrt{s}$ = 8~\TeV\ in 2012.
A CP-odd \textit{Optimal Observable}~\cite{Atwood:1991ka, Davier:1992nw, Diehl:1993br} is employed.
The \textit{Optimal Observable} combines the information from the multi-dimensional phase space in a single quantity calculated from
leading-order matrix elements for VBF production. Hence it does not depend on the decay mode of the Higgs boson.
A direct test of CP invariance is possible measuring the mean value of the CP-odd  \textit{Optimal Observable}.
Moreover, as described in Sect.~2, an ansatz in the framework of an effective field theory is utilised, in which all CP-violating effects corresponding to operators with dimensions up to six
in the couplings between a Higgs boson and an electroweak gauge boson can be described in terms of a single parameter $\dtilde$.
Limits on $\dtilde$ are derived by analysing the shape of spectra of the \textit{Optimal Observable} measured in \Htt\ candidate events that also have two jets tagging VBF production.
The event selection, estimation of background contributions and of systematic 
uncertainties follows the analysis used to establish $4.5\sigma$ evidence for the \Htt\ decay~\cite{HtautauCouplings}. 
Only events selected in the VBF category are analysed, and
only fully leptonic \tll\ or semileptonic \tlhad\ decays of the $\tau$-lepton pair are considered.

The theoretical framework in the context of effective field theories is discussed in Sect.~2
and the methodology of testing CP invariance and the concept of the \textit{Optimal Observable} are introduced in Sect.~3.
After a brief description of the ATLAS detector in Sect.~4,  the simulated samples 
used are summarised in Sect.~5. The experimental analysis is presented in Sect.~6, followed by a description of 
the statistical method used to determine confidence intervals for $\dtilde$ in Sect.~7. The results are discussed in Sect.~8, following which conclusions are given.

%% file: theory.tex
\section{Effective Lagrangian framework}
\label{sec:theory}
The effective Lagrangian considered is the SM Lagrangian augmented by CP-violating operators of mass dimension six,
which can be constructed from the Higgs doublet $\Phi$ and the U(1)$_Y$ and SU(2)$_{I_W,{\rm L}}$ electroweak gauge fields 
$B^\mu$ and $W^{a,\mu}$ ($a$ = 1,2,3), respectively. No CP-conserving dimension-six operators built from 
these fields are taken into account.  All interactions between the Higgs boson and other SM particles (fermions and gluons) are 
assumed to be as predicted in the SM;  i.e.\ the coupling structure in gluon fusion production
 and in the decay into a pair of $\tau$-leptons is considered to be the same as in the SM. 

The effective U(1)$_Y$- and SU(2)$_{I_W,{\rm L}}$-invariant Lagrangian is then given by 
(following Ref.~\cite{Buchmuller:1985jz,Hankele:2006ma}):
\begin{linenomath}
\begin{equation}\label{eq:Leff}
{\cal L}_{\mathrm{eff}} = {\cal L}_{\mathrm{SM}} + \frac{f_{\tilde{B}B}}{\Lambda^2}{\cal O}_{\tilde{B}B} 
                              + \frac{f_{\tilde{W}W}}{\Lambda^2}{\cal O}_{\tilde{W}W} 
                              + \frac{f_{\tilde{B}}}{\Lambda^2}{\cal O}_{\tilde{B}}
\end{equation}
\end{linenomath}

with the three dimension-six operators
\begin{linenomath}\begin{equation}
{\cal O}_{\tilde{B}B} = \Phi^+ \hat{\tilde{B}}_{\mu\nu}  \hat{B}^{\mu\nu} \Phi ~~~~~~
{\cal O}_{\tilde{W}W} = \Phi^+ \hat{\tilde{W}}_{\mu\nu}  \hat{W}^{\mu\nu} \Phi ~~~~~~ 
{\cal O}_{\tilde{B}}  =  (D_\mu \Phi)^+ \hat{\tilde{B}}^{\mu\nu} D_\nu \Phi ~~~~~~.
\end{equation}\end{linenomath}
and three dimensionless Wilson coefficients $f_{\tilde{B}B}$,  $f_{\tilde{W}W}$ and $f_{\tilde{B}}$; $\Lambda$ is the scale of new physics.

Here $D_\mu$ denotes the covariant derivative $D_\mu = \partial_\mu +\frac{\mathrm{i}}{2}g^\prime B_\mu 
+ \mathrm{i} g \frac{\sigma_a}{2}  W_\mu^a$,
$\hat{V}_{\mu\nu}$  ($V = B, W^a$) the field-strength tensors and  
$\tilde{V}_{\mu\nu} = \frac{1}{2} \epsilon_{\mu \nu \rho \sigma}{V}^{\rho\sigma}$
the dual field-strength tensors, with 
$\hat{B}_{\mu\nu} + \hat{W}_{\mu\nu} = \mathrm{i} \frac{g^\prime}{2} B_{\mu\nu} + \mathrm{i} \frac{g}{2} \sigma^a W^a_{\mu\nu}$.

The last operator ${\cal O}_{\tilde{B}}$ contributes to the CP-violating charged triple gauge-boson couplings 
$\tilde{\kappa}_{\gamma}$ and $\tilde{\kappa}_{Z}$ via the relation 
$\tilde{\kappa}_{\gamma} = - \cot^2 \theta_W \tilde{\kappa}_{Z} = \frac{m_W^2}{2\Lambda^2} f_{\tilde{B}}$.  
These CP-violating charged triple gauge boson couplings are constrained by the 
LEP experiments~\cite{Abbiendi:2000ei,Schael:2004tq,Abdallah:2008sf} and the contribution from 
${\cal O}_{\tilde{B}}$ is neglected in the following; i.e.\ only
contributions from  ${\cal O}_{\tilde{B}B}$ and ${\cal O}_{\tilde{W}W}$ are taken into account.

After electroweak symmetry breaking in the unitary gauge the effective Lagrangian in the mass basis of 
Higgs boson $H$, photon $A$ and weak gauge bosons $Z$ and $W^\pm$ can be written, e.g.\ as in Ref.~\cite{Achard:2004kn}:
\begin{linenomath}
\begin{equation}
{\cal L}_{\mathrm{eff}} = {\cal L}_{\mathrm{SM}} + \tilde{g}_{HAA} H \tilde{A}_{\mu\nu}{A}^{\mu\nu} +
                \tilde{g}_{HAZ} H \tilde{A}_{\mu\nu}{Z}^{\mu\nu} +
                \tilde{g}_{HZZ} H \tilde{Z}_{\mu\nu}{Z}^{\mu\nu} +
                \tilde{g}_{HWW} H \tilde{W}^+_{\mu\nu}{W}^{-\mu\nu} \, . 
\end{equation}
\end{linenomath}
Only two of the four couplings $\tilde{g}_{HVV}$ ($V=W^\pm,Z,\gamma$) are independent due to constraints imposed 
by U(1)$_Y$ and SU(2)$_{I_W,{\rm L}}$ invariance. They can  be expressed in terms of two dimensionless couplings 
$\tilde{d}$ and $\tilde{d}_B$ as:
\begin{linenomath}
\begin{eqnarray}\label{eq:translations}
\tilde{g}_{HAA}  =   \frac{g}{2 m_W} (\tilde{d} \sin^2 \theta_W + \tilde{d}_B \cos^2 \theta_W) & & 
\tilde{g}_{HAZ}  =   \frac{g}{2 m_W} \sin 2\theta_W (\tilde{d} -\tilde{d}_B)\\
\tilde{g}_{HZZ}  =  \frac{g}{2 m_W} (\tilde{d} \cos^2 \theta_W + \tilde{d}_B \sin^2 \theta_W) & & 
\tilde{g}_{HWW}  =   \frac{g}{m_W} \tilde{d} \, .
\end{eqnarray}\end{linenomath}
Hence in general $WW$, $ZZ$, $Z\gamma$ and $\gamma\gamma$ fusion contribute to VBF production.
The relations between $\tilde{d}$ and $f_{\tilde{W}W}$, and $\tilde{d}_B$ and  $f_{\tilde{B}B}$ are given by:
\begin{linenomath}\begin{equation}
\tilde{d}   = - \frac{m_W^2}{\Lambda^2}                 f_{\tilde{W}W} ~~~~~~~~~~~~  
\tilde{d}_B = - \frac{m_W^2}{\Lambda^2} \tan^2 \theta_W  f_{\tilde{B}B} \, .
\end{equation}\end{linenomath}
As the different contributions from the various electroweak gauge-boson fusion processes cannot
be distinguished experimentally with the current available dataset, the arbitrary choice $\tilde{d} = \tilde{d}_B$ is adopted.
This yields the following relation for the $\tilde{g}_{HVV}$:
\begin{linenomath}\begin{equation}
\tilde{g}_{HAA} = \tilde{g}_{HZZ} = \frac{1}{2} \tilde{g}_{HWW} = \frac{g}{2 m_W} \tilde{d}~~~~   
{\mathrm{and}} ~~~~~~ \tilde{g}_{HAZ} = 0 \, .
\end{equation}\end{linenomath}

The parameter $\tilde{d}$ is related to the parameter $\hat{\kappa}_W= \tilde{\kappa}_W/ \kappa_{\mathrm{SM}} \tan \alpha$ used in the investigation
of CP properties in the decay $\HWW$~\cite{Aad:2015mxa}
via $\tilde{d} = - \hat{\kappa}_W$.
The choice $\tilde{d} = \tilde{d}_B$ yields $\hat{\kappa}_W = \hat{\kappa}_Z$ 
as assumed in the combination of the $\HWW$ and $H\rightarrow ZZ$ decay 
analyses~\cite{Aad:2015mxa}.

The effective Lagrangian yields the following Lorentz structure for each vertex in the Higgs
bosons coupling to two identical or charge-conjugated electroweak gauge bosons $HV(p_1)V(p_2)$ 
($V=W^{\pm},Z,\gamma$), with $p_{1,2}$ denoting the momenta of the gauge bosons:
\begin{linenomath}\begin{equation}
 T^{\mu\nu} (p_1,p_2) = \sum_{V=W^{\pm},Z} \frac{2m^{2}_{V}}{v} g^{\mu\nu} +  \sum_{V=W^{\pm},Z,\gamma} 
\frac{2g}{m_W} \tilde{d} ~ \varepsilon^{\mu \nu \rho \sigma} p_{1\rho} p_{2\sigma} \, .
\end{equation}\end{linenomath}
The first terms ($\propto g^{\mu\nu}$) are CP-even and describe the SM coupling structure, while
the second terms ($\propto \varepsilon^{\mu \nu \rho \sigma} p_{1\rho} p_{2\sigma}$) 
are CP-odd and arise from the CP-odd dimension-six operators. The choice
$\tilde{d} = \tilde{d}_B$ gives the same coefficients multiplying the CP-odd structure
for $HW^+W^-$, $HZZ$ and $H\gamma\gamma$ vertices and a vanishing coupling for the $HZ\gamma$ vertex.

The matrix element $\mathcal{M}$ for VBF production is the sum of a CP-even contribution $\mathcal{M}_{\textrm{SM}}$
from the SM and a CP-odd contribution $\mathcal{M}_{\textrm{CP-odd}}$ from the dimension-six operators considered:
\begin{linenomath}\begin{equation} 
\mathcal{M} =\mathcal{M}_{\textrm{SM}}+\dtilde\cdot\mathcal{M}_{\textrm{CP-odd}}.
\end{equation}\end{linenomath}
The differential cross section or squared matrix element has three contributions:
\begin{linenomath}\begin{equation} \label{eq:MEsquared}
|\mathcal{M}|^{2}=|\mathcal{M}_{\textrm{SM}}|^{2}+ 
\dtilde\cdot2 \operatorname{Re}(\mathcal{M}_{\textrm{SM}}^{*}\mathcal{M}_{\textrm{CP-odd}}) +  \dtilde^{2}\cdot|\mathcal{M}_{\textrm{CP-odd}}|^{2} \, .
\end{equation}\end{linenomath}
The first term $|\mathcal{M}_{\textrm{SM}}|^{2}$ and third term 
$\dtilde^{2}\cdot|\mathcal{M}_{\textrm{CP-odd}}|^{2}$ are both CP-even and hence do not yield a source 
of CP violation. The second term 
$\dtilde\cdot2 \operatorname{Re}(\mathcal{M}_{\textrm{SM}}^{*}\mathcal{M}_{\textrm{CP-odd}})$, stemming from the interference of the 
two contributions to the matrix element, is CP-odd and is a possible new source of CP violation in the Higgs sector.
The interference term integrated over a CP-symmetric part of phase space vanishes and therefore does 
not contribute to the total cross section and observed event yield after applying CP-symmetric selection criteria.
The third term increases the total cross section by an amount quadratic in $\dtilde$, but this is not 
exploited in the analysis presented here.

%% file: method.tex
\section{Test of CP invariance and \textit{Optimal Observable}}
\label{sec:method}
Tests of CP invariance can be performed in a completely model-independent way by measuring the mean
value of a CP-odd observable $\langle \mathcal{O}_{\textrm{CP}} \rangle$. If CP invariance holds, the mean value has 
to vanish $\langle \mathcal{O}_{\textrm{CP}} \rangle = 0$. An observation of a non-vanishing mean value would be 
a clear sign of CP violation. A simple CP-odd observable for Higgs boson production in VBF, 
the ``signed'' difference in the azimuthal angle  between the two tagging jets $\Delta \phi_{jj}$, 
was suggested in Ref.~\cite{Hankele:2006ma} and is formally defined as:
\begin{linenomath}\begin{equation} 
\label{eq:signdphi}
\epsilon_{\mu \nu \rho \sigma} b_+^\mu p_+^\nu  b_-^\rho p_-^\sigma 
= 2 p_{\rm T+} p_{\rm T-} \sin(\phi_+ - \phi_-) = 2 p_{\rm T+} p_{\rm T-}  \sin\Delta \phi_{jj} \, .  
\end{equation}\end{linenomath}
Here $b_+^\mu$ and $b_-^\mu$ denote the normalised four-momenta of the two proton beams,
circulating clockwise and anti-clockwise, and $p_+^\mu$ ($\phi_+$) and $p_-^\mu$ ($\phi_-$) 
denote the four-momenta (azimuthal angles) of the two tagging jets, 
where $p_+$ ($p_-$) points into the same detector hemisphere as $b_+^\mu$  ($b_-^\mu$).
This ordering of the tagging jets by hemispheres removes the sign ambiguity 
in the standard definition of $\Delta \phi_{jj}$.

The final state consisting of the Higgs boson and the two tagging jets can 
be characterised by seven phase-space variables while assuming the mass of  
the Higgs boson, neglecting jet masses and exploiting momentum conservation
in the plane transverse to the beam line. The concept of the 
\textit{Optimal Observable} combines the information
of the high-dimensional phase space in a single observable, which can be 
shown to have the highest sensitivity for small values of the parameter of 
interest and neglects contributions proportional to $\dtilde^2$ in the matrix element.
The method was first suggested for the estimation of a single parameter 
using the mean value only~\cite{Atwood:1991ka} and via  
a maximum-likelihood fit to the full distribution~\cite{Davier:1992nw}
using the so-called \textit{Optimal Observable} of first order.
The extension to several parameters and also exploiting the matrix-element  
contributions quadratic in the parameters by adding an 
\textit{Optimal Observable} of second order was introduced in Refs.~\cite{Diehl:1993br,Diehl:1997ft,Diehl:2002nj}. The technique has been applied in 
various experimental analyses, e.g.\ Refs.~\cite{Buskulic:1993vk,Abreu:1995ku,Acciarri:1998vg,Abbiendi:2001km,OOOPAL,Ackerstaff:1996gy,Barate:1997ts,Abreu:1999vv,Acciarri:1999kn,Abbiendi:1998bg,Schumacher:2001ax,Aad:2015mxa}.

\begin{figure}[h!]
  \centering
   \includegraphics[width=0.49\textwidth]{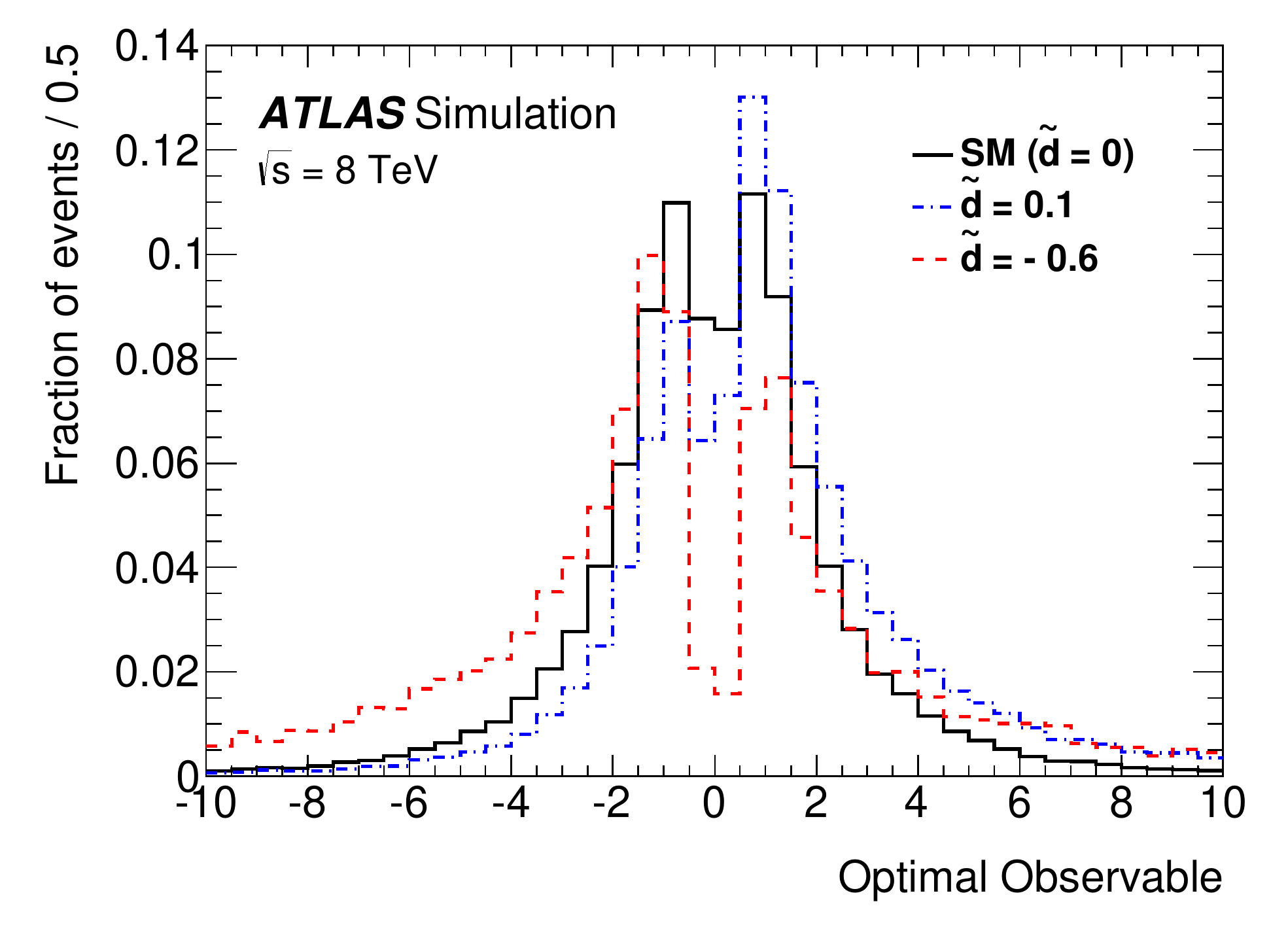}  
   \caption{Distribution of the \textit{Optimal Observable} at parton-level for two arbitrary $\dtilde$ values. The SM sample was generated using 
\textsc{MadGraph5\_aMC@NLO}~\cite{aMCatNLO} (see Sect.~\ref{sec:samples}) at leading order, and then reweighted to different $\dtilde$ values. Events are chosen such that there are at least two outgoing partons with $\pt > 25 \GeV$, $|\eta| < 4.5 $, large invariant mass ($m(p_1,p_2) > 500 \GeV $) and large pseudorapidity gap ($\Delta \eta(p_1,p_2) > 2.8$ ).} 
  \label{fig:ooGenLevel}
\end{figure}

The analysis presented here uses only the first-order \textit{Optimal Observable} 
{$\mathcal{\,OO\,}$} (called \textit{Optimal Observable} below) 
for the measurement of $\dtilde$ via maximum-likelihood fit to the full distribution.
It is defined as the ratio of the interference term in the matrix element to the 
SM contribution:
\begin{linenomath}\begin{equation}
\label{eq:OO} 
\mathcal{OO} =\frac{2 \operatorname{Re}(\mathcal{M}_{\textrm{SM}}^{*}\mathcal{M}_{\textrm{CP-odd}})}{|\mathcal{M}_{\textrm{SM}}|^{2}} \, . 
\end{equation}\end{linenomath}
Figure~\ref{fig:ooGenLevel} shows the distribution of the \textit{Optimal Observable}, 
at parton level both for the SM case and for two non-zero  $\dtilde$ values, which 
introduce an asymmetry into the distribution and yield a non-vanishing mean value.

The values of the leading-order matrix elements needed for the calculation of the \textit{Optimal Observable} are 
extracted from HAWK~\cite{Ciccolini:2007jr,Ciccolini:2007ec,Denner:2014cla}. The evaluation requires
the four-momenta of the Higgs boson and the two tagging jets. 
The momentum fraction $x_{1}$ ($x_{2}$) of the initial-state parton from the proton moving in the positive (negative)
 $z$-direction can be derived 
by exploiting energy--momentum conservation from the Higgs boson and tagging jet four-momenta as:
\begin{linenomath}\begin{equation}
x_{1/2}^{\textrm{reco}} = \frac{m_{Hjj}}{\sqrt{s}}\mathrm{e}^{\pm y_{Hjj}} ~~~~~~ 
\end{equation}\end{linenomath}
where $m_{\textrm{Hjj}}$ ($y_{\textrm{Hjj}}$) is the invariant mass (rapidity) obtained from the vectorially summed four-momenta of the 
tagging jets and the Higgs boson.
Since the flavour of the initial- and final-state partons cannot be determined experimentally,
the sum over all possible flavour configurations $ij \rightarrow kl H$
weighted by the CT10 leading-order parton distribution functions (PDFs)~\cite{Lai:2010vv}
is calculated separately for the matrix elements in the numerator and denominator:
\begin{linenomath}\begin{eqnarray}
2 \operatorname{Re}(\mathcal{M}_{\textrm{SM}}^{*}\mathcal{M}_{\textrm{CP-odd}}) &=& \sum_{i,j,k,l} 
f_i(x_1) f_j(x_2) 2 \operatorname{Re}((\mathcal{M}_{\textrm{SM}}^{ij \rightarrow kl H})^*\mathcal{M}_{\textrm{CP-odd}}^{ij \rightarrow kl H})\\
|\mathcal{M}_{\mathrm{SM}}|^2 &=&  \sum_{i,j,k,l} f_i(x_1) f_j(x_2) |\mathcal{M}_{\mathrm{SM}}^{ij \rightarrow kl H}|^2\, . 
\end{eqnarray}\end{linenomath}

%% file: atlas.tex
\section{The ATLAS detector}
\label{sec:atlas}
The ATLAS detector~\cite{DetectorPaper:2008} is a multi-purpose detector with a
cylindrical geometry.\footnote{ATLAS uses a right-handed coordinate system with its origin at the nominal interaction point (IP) in the centre of the detector and the $z$-axis along the beam pipe. The $x$-axis points from the IP to the centre of the LHC ring, and the $y$-axis points upward. Cylindrical coordinates $(r,\phi)$ are used in the transverse plane, $\phi$ being the azimuthal angle around the $z$-axis. The pseudorapidity is defined in terms of the polar angle $\theta$ as $\eta=-\ln\tan(\theta/2)$.} It comprises an inner detector (ID) surrounded by a thin
superconducting solenoid, a calorimeter system and an extensive muon
spectrometer  in a toroidal magnetic field. The ID tracking system
consists of a silicon pixel detector, a silicon microstrip detector, and
a transition radiation tracker.  It provides precise position and
momentum measurements for charged particles and allows efficient identification
of jets containing $b$-hadrons ($b$-jets) in the pseudorapidity  range $|\eta|<2.5$. The ID
is immersed in a 2 T axial magnetic field and is surrounded by high-granularity
 lead/liquid-argon sampling electromagnetic calorimeters which
cover the pseudorapidity range $ |\eta|< 3.2$. A steel/scintillator tile
calorimeter provides hadronic energy measurements in the central pseudorapidity
range ($ |\eta|< 1.7$). In the forward regions ($1.5 < |\eta| < 4.9$), the
system is complemented by two end-cap calorimeters using liquid argon as active material
and copper or tungsten as absorbers. The muon spectrometer  surrounds the
calorimeters and consists of three large superconducting eight-coil toroids, a
system of tracking chambers, and detectors for triggering.  The deflection of
muons is measured in the region $|\eta|< 2.7$ by three layers of precision drift
tubes, and cathode strip chambers in the innermost layer for $|\eta| > 2.0$.
The trigger chambers consist of resistive plate chambers in the barrel ($|\eta|
< 1.05$) and thin-gap chambers in the end-cap regions ($1.05<|\eta|<2.4$).

A three-level trigger system~\cite{TriggerPapaer:2012} is used to select events.
A hardware-based Level-1 trigger uses a subset of detector information to reduce the event
rate to $75$~kHz or less. The rate of accepted events is then reduced to about
$400$~Hz by two software-based trigger levels, named Level-2 and the Event Filter.

%% file: samples.tex
\section{Simulated samples}
\label{sec:samples}

Background and signal events are simulated using various Monte
Carlo (MC) event generators, as summarised in Table~\ref{tab:MCGenerator}. The
generators used for the simulation of the hard-scattering process and the model
used for the simulation of the parton shower, hadronisation and underlying-event activity are listed. In addition, the cross-section values to
which the simulation is normalised and the perturbative order in QCD of the
respective calculations are provided.

\begin{table}
  \begin{center}
  \small
    \begin{tabular}{|l|l|lrl|}
      \hline
      \multirow{2}{*}{Signal}   & \multirow{2}{*}{MC generator}  & \multicolumn{3}{l|}{$\sigma \times {\cal B}$ [pb]} \\
                                 &                                  &  \multicolumn{3}{l|}{$\sqrt{s}=8~\TeV$}  \\
      \hline \hline
       VBF, $\Htt$     &  \textsc{Powheg-Box}~\cite{Nason:2004rx,Frixione:2007vw,Alioli:2010xd,POWHEGVBF} \textsc{Pythia8}~\cite{Sjostrand:2007gs} & 0.100  & (N)NLO    & \cite{Ciccolini:2007jr, Ciccolini:2007ec, Arnold:2008rz,Bolzoni:2010xr,YellowReportIII}\\
        VBF, $\HWW$             & same as for $\Htt$ signal               & 0.34 &  (N)NLO    & \cite{Ciccolini:2007jr, Ciccolini:2007ec, Arnold:2008rz,Bolzoni:2010xr,YellowReportIII} \\
      \hline \hline
      \multirow{2}{*}{Background} & \multirow{2}{*}{MC generator} & \multicolumn{3}{l|}{$\sigma \times {\cal B}$ [pb]}\\
                 &              &  \multicolumn{3}{l|}{$\sqrt{s}=8~\TeV$}            \\
      \hline \hline
      $W (\to \ell \nu$), ($\ell = e, \mu, \tau)$                  & \textsc{Alpgen}~\cite{alpgen}+\textsc{Pythia8}   & 36800  & NNLO & \cite{PhysRevLett.103.082001,PhysRevLett.98.222002}\\
      $Z/\gamma^{*}(\to \ell\ell)$,& \multirow{2}{*}{\textsc{Alpgen+Pythia8}}   & \multirow{2}{*}{3910} & \multirow{2}{*}{NNLO} & \multirow{2}{*}{\cite{PhysRevLett.103.082001,PhysRevLett.98.222002}}\\
      60~\GeV$<m_{\ell\ell}<2$~\TeV   & & & &\\
      $Z/\gamma^{*}(\to \ell\ell)$,   & \multirow{2}{*}{\textsc{Alpgen+Herwig}~\cite{herwig}}       & \multirow{2}{*}{13000}  & \multirow{2}{*}{NNLO} & \multirow{2}{*}{\cite{PhysRevLett.103.082001,PhysRevLett.98.222002}}\\
      10~\GeV$<m_{\ell\ell}<60$~\GeV  & & & &\\
      VBF $Z/\gamma^{*}$($\to\ell\ell$)                           & \textsc{Sherpa}~\cite{sherpa}              &  1.1  & LO & \cite{sherpa}\\
      $t\bar{t}$                         & \textsc{Powheg-Box + Pythia8} & 253$^{\dagger}$  & NNLO+NNLL & \cite{ttbar:xsec1,ttbar:xsec2,ttbar:xsec3,ttbar:xsec4,ttbar:xsec5,ttbar:xsec6}\\
      Single top : $Wt$                  & \textsc{Powheg-Box + Pythia8}                                          & 22$^{\dagger}$ & NNLO & \cite{singletop:Wt}\\
      Single top : $s$-channel           & \textsc{Powheg-Box + Pythia8}                                   & 5.6$^{\dagger}$ & NNLO & \cite{singletop:s-ch}\\
      Single top : $t$-channel           & AcerMC~\cite{Kersevan:2004yg}+\textsc{Pythia6}~\cite{Pythia}             & 87.8$^{\dagger}$ & NNLO & \cite{singletop:t-ch}\\
      $q\bar{q} \to WW$                  & \textsc{Alpgen+Herwig}                                           & 54$^{\dagger}$ & NLO & \cite{MCFMVV}\\
      $gg \to WW $                       & \textsc{gg2WW}~\cite{Binoth:2006mf}+\textsc{Herwig}                       & 1.4$^{\dagger}$ & NLO & \cite{Binoth:2006mf}\\
      $WZ,ZZ$                            & \textsc{Herwig}                                                  & 30$^{\dagger}$ & NLO & \cite{MCFMVV}\\
      ggF, $\Htt$   & \textsc{HJ MINLO}~\cite{POWHEGHJ,MINLO} + \textsc{Pythia8}  & 1.22   & NNLO+NNLL & \cite{Djouadi:1991tka, Dawson:1990zj, Spira:1995rr, Harlander:2002wh, Anastasiou:2002yz, Ravindran:2003um,YellowReportIII}\\
      ggF, $\HWW$                         & \textsc{Powheg-Box}~\cite{Bagnaschi:2011tu} + \textsc{Pythia8}  & 4.16  & NNLO+NNLL & \cite{Djouadi:1991tka, Dawson:1990zj, Spira:1995rr, Harlander:2002wh, Anastasiou:2002yz, Ravindran:2003um,YellowReportIII} \\
     \hline
    \end{tabular}
\caption{
MC event generators used to model the signal and the background processes at
$\sqrt{s}=8\TeV$. All Higgs boson events are generated assuming $m_{H} = 125 \GeV$.
The cross sections times branching fractions ($\sigma \times{\cal B}$) used for the normalisation of some processes (many of these are
subsequently normalised to data) are included in the last column together with
the perturbative order of the QCD calculation.  For the signal processes the
$\Htt$ and $\HWW$ SM branching ratios are included, and for the $W$ and $Z/\gamma^{*}$
background processes the branching ratios for leptonic decays ($\ell = e, \mu,\tau$) 
of the bosons are included. For all other background processes,
inclusive cross sections are quoted (marked with a $\dagger$).}
\label{tab:MCGenerator}
  \end{center}
\end{table}

All the background samples used in this analysis are the same as those employed in Ref.~\cite{HtautauCouplings}, except the ones used to simulate events with the Higgs boson produced via gluon fusion and decaying into the $\tau\tau$ final state.
The Higgs-plus-one-jet process is simulated at NLO accuracy in QCD with \textsc{Powheg-Box}~\cite{Nason:2004rx,Frixione:2007vw,Alioli:2010xd,POWHEGHJ}, 
with the \textsc{MINLO} feature ~\cite{MINLO} applied to include Higgs-plus-zero-jet events at NLO accuracy.  
This sample is referred to as HJ \textsc{MINLO}. 
The \textsc{Powheg-Box} event generator is interfaced to \textsc{Pythia8}~\cite{Sjostrand:2007gs}, and 
the \textsc{CT10}~\cite{Lai:2010vv} parameterisation of the PDFs is used. 
Higgs boson events produced via gluon fusion and decaying into the $W^{+}W^{-}$ final state, which are a small component of the background, are simulated, as in Ref.~\cite{HtautauCouplings}, with \textsc{Powheg}~\cite{Nason:2004rx,Frixione:2007vw,Alioli:2010xd,Bagnaschi:2011tu} interfaced to \textsc{Pythia8}~\cite{Sjostrand:2007gs}. 
For these simulated events, the shape of the generated \pt\ distribution is matched to a NNLO+NNLL calculation \textsc{HRes2.1}~\cite{hres_1}~\cite{hres_2} in the inclusive phase space. Simultaneously, for events with two or more jets, the Higgs boson \pt\ spectrum is reweighted to match the \textsc{MINLO HJJ} predictions~\cite{minlo_hjj}.
The overall normalisation of the gluon fusion process (ggF)  is taken from a calculation at
next-to-next-to-leading order (NNLO)~\cite{Djouadi:1991tka, Dawson:1990zj,Spira:1995rr, Harlander:2002wh, Anastasiou:2002yz, Ravindran:2003um}
in QCD, including soft-gluon resummation up to next-to-next-to-leading
logarithm terms (NNLL)~\cite{Catani2003}. Next-to-leading-order (NLO) electroweak
(EW) corrections are also included~\cite{Aglietti:2004nj,Actis:2008ug}. 
Higgs boson events produced via VBF, with SM couplings, are also simulated with \textsc{Powheg} interfaced with \textsc{Pythia8} 
(see Table~\ref{tab:MCGenerator} and Ref.~\cite{HtautauCouplings}).

Production by VBF is normalised to a cross section calculated with full NLO
QCD and EW corrections~\cite{Ciccolini:2007jr, Ciccolini:2007ec, Arnold:2008rz}
with an approximate NNLO QCD correction applied~\cite{Bolzoni:2010xr}.
The NLO EW corrections for VBF production depend on the $\pT$ of the
Higgs boson, and vary from a few percent at low $\pT$ to $\sim 20\%$ at
$\pT$~=~300~\GeV~\cite{Dittmaier:12013084}. The $\pT$ spectrum of the
VBF-produced Higgs boson is therefore reweighted, based on the difference between the
\textsc{Powheg-Box}+\textsc{Pythia} calculation and the
\textsc{Hawk}~\cite{Ciccolini:2007jr,Ciccolini:2007ec,Denner:2014cla} calculation which
includes these corrections.

In the case of VBF-produced Higgs boson events in the presence of anomalous couplings in the $HVV$ vertex, the simulated samples are 
obtained by applying a matrix element (ME) reweighting method to the VBF SM signal sample.
The weight is defined as the ratio of the squared ME value for the VBF process associated with a specific amount of CP mixing
(measured in terms of $\dtilde$) to the SM one. The inputs needed for the ME evaluation are  the flavour of the incoming partons, 
the four-momenta and the flavour of the two or three final-state  partons and the four-momentum of the Higgs boson.
The Bjorken $x$ values of the initial-state partons can be calculated from energy--momentum conservation.  
The leading-order ME from \textsc{HAWK}~\cite{Ciccolini:2007jr,Ciccolini:2007ec,Denner:2014cla} is used for the $2\to2+H$ or $2\to3+H$ process separately.
This reweighting procedure is validated against samples generated with \textsc{MadGraph5\_aMC@NLO}~\cite{aMCatNLO}.  
As described in Ref.~\cite{HiggsCharacterisationModel}, \textsc{MadGraph5\_aMC@NLO} can simulate VBF production with anomalous couplings 
at next-to-leading order. The reweighting procedure proves to be a good approximation to a full next-to-Leading description of the BSM process.

In the case of the \HWW\ sample, if CP violation exists in the $HVV$ coupling, it would affect both the VBF production and the $HWW$ decay vertex. It was verified that the shape of the 
\textit{Optimal Observable} distribution is independent of any possible CP violation in the \HWW\ decay vertex and that it is identical for \HWW\ and \Htt\ decays.
Hence the same reweighting is applied for VBF-produced events with \HWW\ and \Htt\ decays.

For all  samples, a full simulation of the ATLAS detector
response~\cite{Aad:2010ah} using the \textsc{Geant4}
program~\cite{Agostinelli:2002hh} was performed. In addition, multiple simultaneous minimum-bias interactions
 are simulated using the
\textsc{AU2}~\cite{atlasmctunes} parameter tuning of \textsc{Pythia8}. 
They are overlaid on the simulated signal and background events according to the luminosity
profile of the recorded data. The contributions from these pile-up interactions
are simulated both within the same bunch crossing as the hard-scattering
process and in neighbouring bunch crossings. Finally, the resulting simulated
events are processed through the same reconstruction programs as the data.

%% file: analysis.tex
\section{Analysis}
\label{sec:analysis}

After data quality requirements, the integrated luminosity of the \mbox{$\sqrt{s}=8\TeV$} dataset used is 20.3~\fbs.
The triggers, event selection, estimation of background contributions and systematic uncertainties closely follow the analysis in Ref.~\cite{HtautauCouplings}.
In the following a short description of the analysis strategy is given; more details are given in that reference.

Depending on the reconstructed decay modes of the two $\tau$ leptons (leptonic or hadronic), events are separated into 
the dileptonic (\tll) and semileptonic (\tlhad) channels. Following a channel-specific preselection, a VBF region is selected by requiring at least two jets with $\pt^{j_1}$ $>$ 40~\GeV\ (50~\GeV) and $\pt^{j_2}>30$~\GeV\ and a pseudorapidity separation $\Delta\eta(j_1,j_2) > 2.2$ (3.0) in the \tll\ (\tlhad) channel. Events with $b$-tagged jets are removed to suppress top-quark backgrounds.

\begin{figure}[h!]
  \centering
   \subfloat[]{\includegraphics[width=0.49\textwidth]{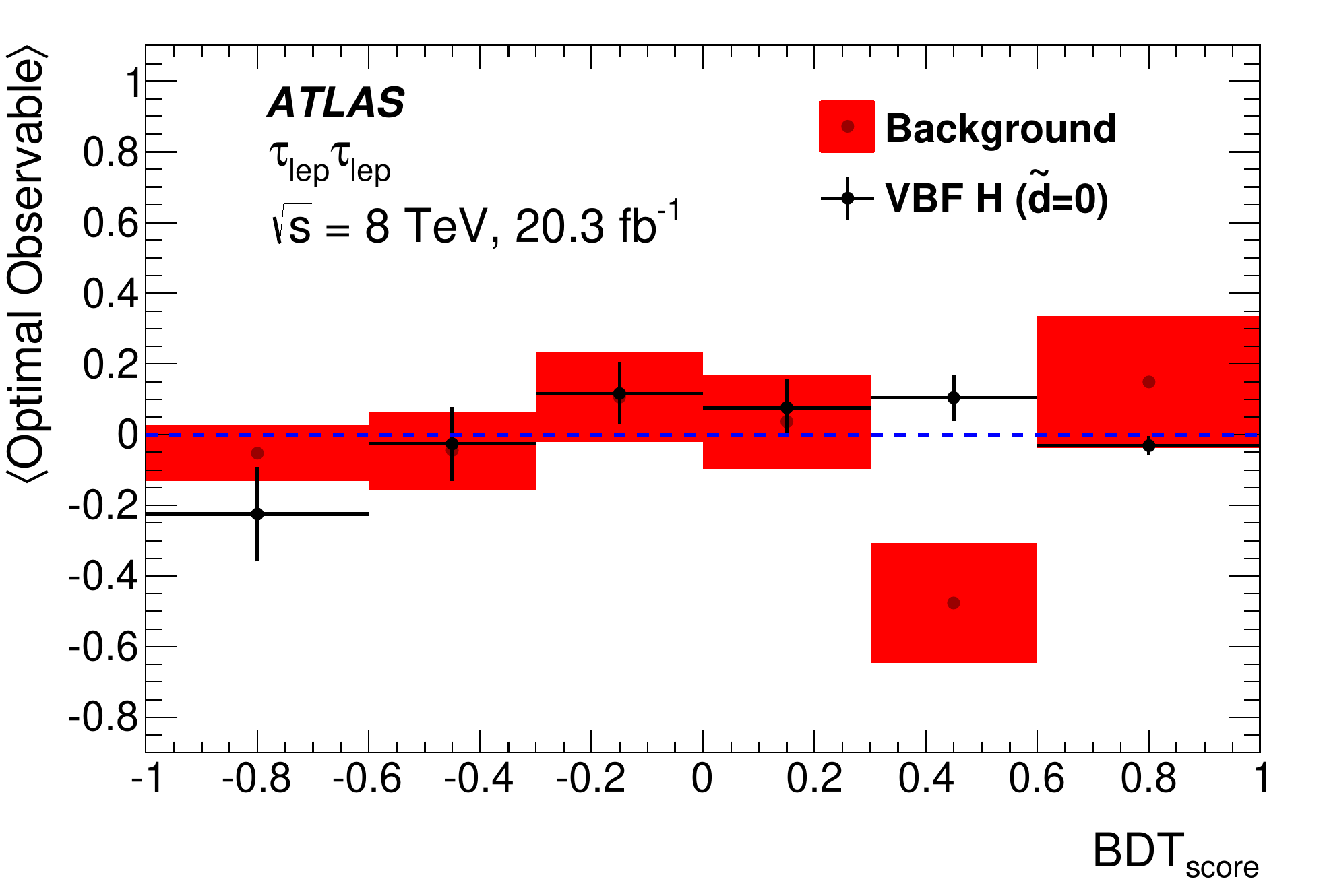} }
   \subfloat[]{\includegraphics[width=0.49\textwidth]{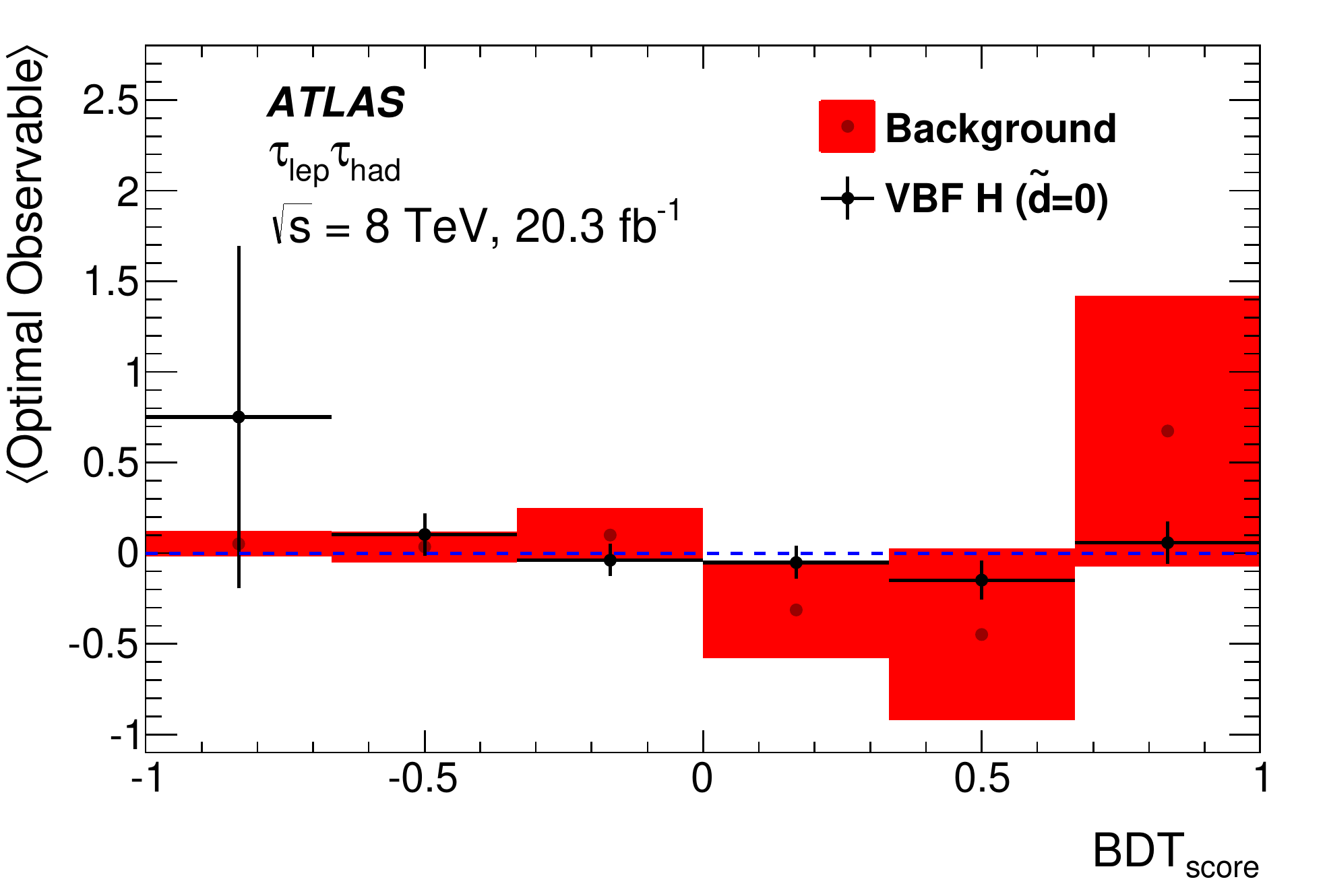}  }
   \caption{Mean of the \textit{Optimal Observable} as a function of the \bdtscore\ for the SM signal (black dots with error bars) and for the sum of all background processes (filled red area), for the (a) \tll\ and (b) \tlhad\ 
   channel. The signal and background model is in agreement with the hypothesis of no bias from the BDT score.} 
  \label{fig:bdtmean}
\end{figure}

Inside the VBF region, boosted decision trees (BDT)\footnote{The same BDTs trained in the context of the analysis in Ref.~\cite{HtautauCouplings} are used here, unchanged.} are  utilised  for separating Higgs boson events produced via VBF from the background (including other Higgs boson production modes). The final signal region in each channel is defined by the events with a \bdtscore\ value above a  threshold of 0.68 for \tll\ and 0.3 for \tlhad. The efficiency of this selection, with respect to the full VBF region, is 49\% (51\%) for the signal and 3.6\% (2.1\%) for the sum of background processes for the \tll\ (\tlhad) channel. A non-negligible number of events from VBF-produced \HWW\ events survive the $\tll$ selection: they amount to 17\% of the overall VBF signal in the signal region. Their contribution is entirely negligible in the \tlhad\ selection. 
Inside each signal region, the \textit{Optimal Observable} is then used as the variable with which to probe for CP violation.  
The \bdtscore\ does not affect the mean of the \textit{Optimal Observable}, as can be seen in Fig.~\ref{fig:bdtmean}.

\begin{figure}[h!]
  \centering
   \subfloat[]{\includegraphics[width=0.49\textwidth]{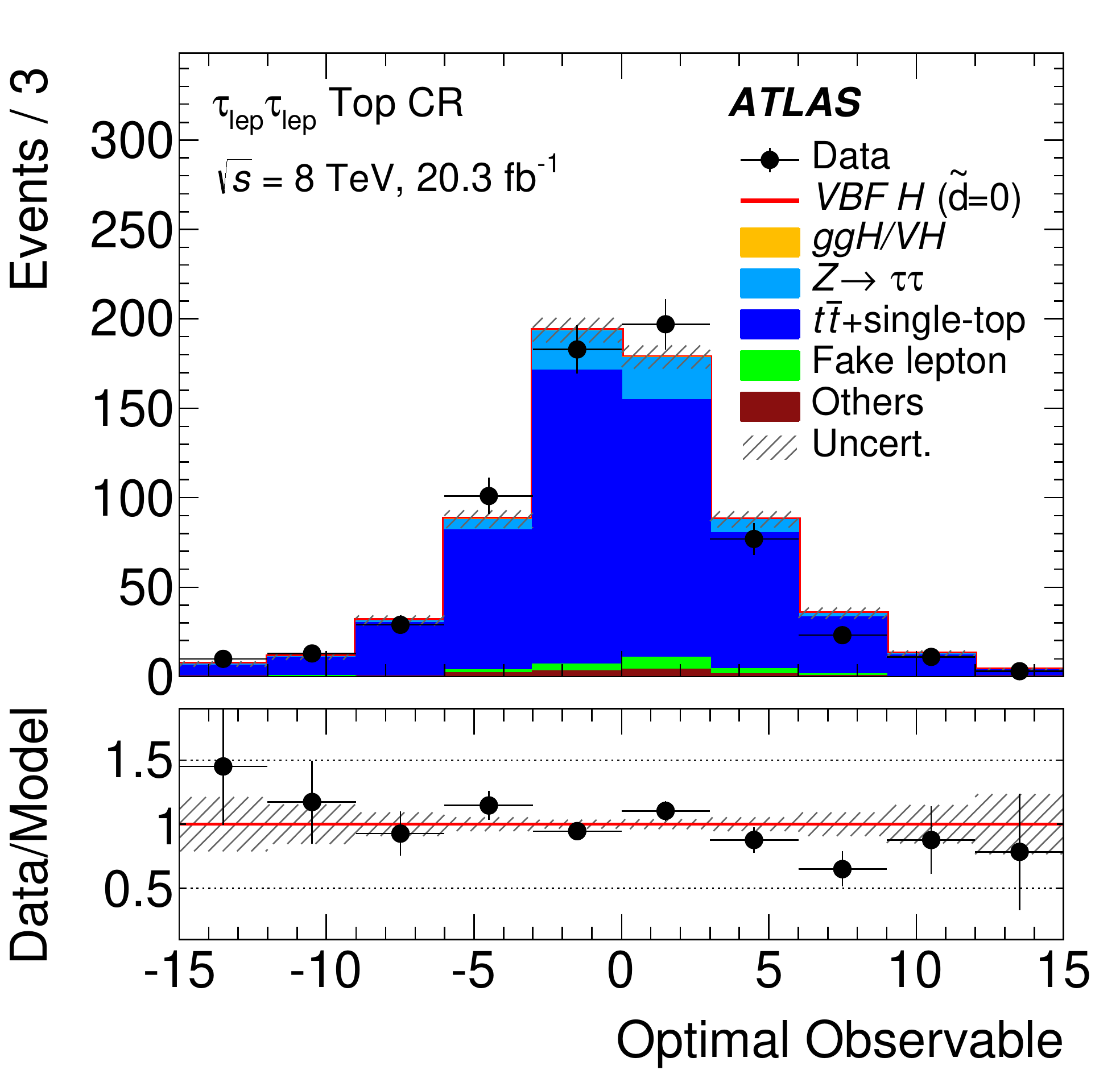}  }
   \subfloat[]{\includegraphics[width=0.49\textwidth]{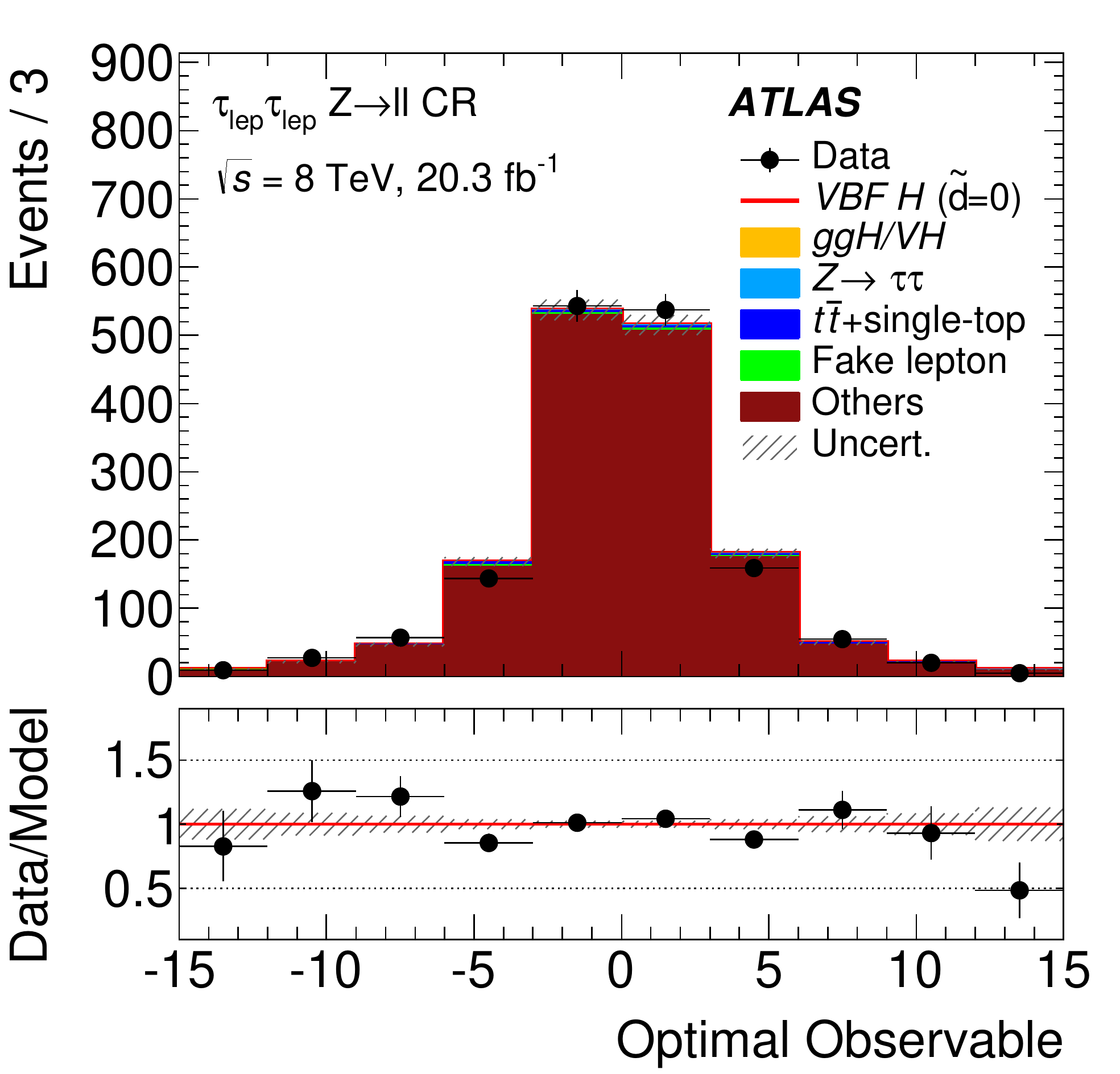} } \\
   \subfloat[]{\includegraphics[width=0.49\textwidth]{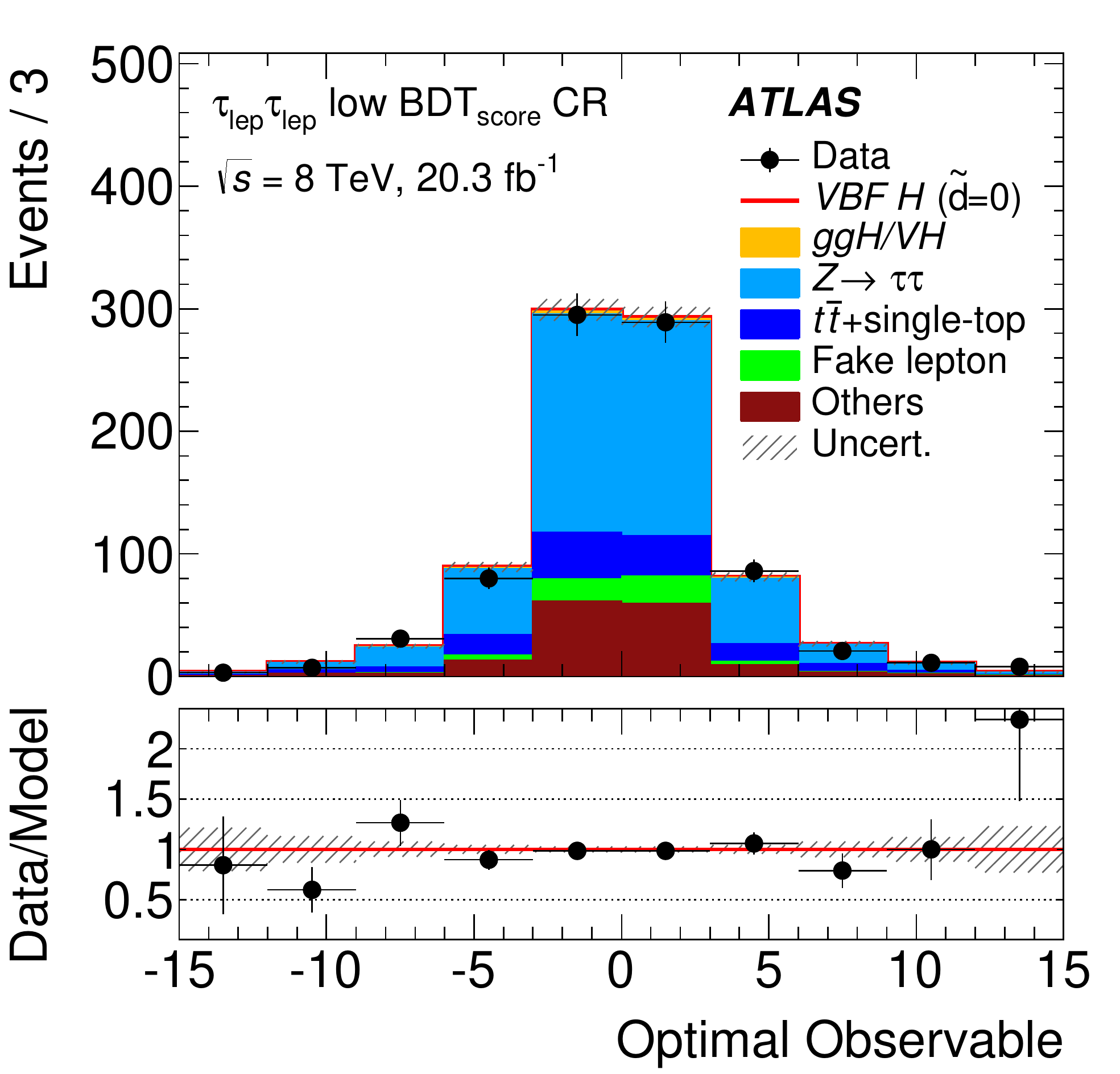}  }
   \caption{Distributions of the \textit{Optimal Observable} for the \tll\  channel  in the (a) top-quark control region (CR), (b)  \Zll\ CR, and (c) low-\bdtscore\ CR. The CR definitions are given in the text. These figures use background predictions before the global fit defined in Sect.~\ref{sec:fit}. The ``Other'' backgrounds include diboson and \Zll. Only statistical uncertainties are shown. 
} 
  \label{fig:crsll}
\end{figure}

\begin{figure}[h!]
  \centering
   \subfloat[]{\includegraphics[width=0.49\textwidth]{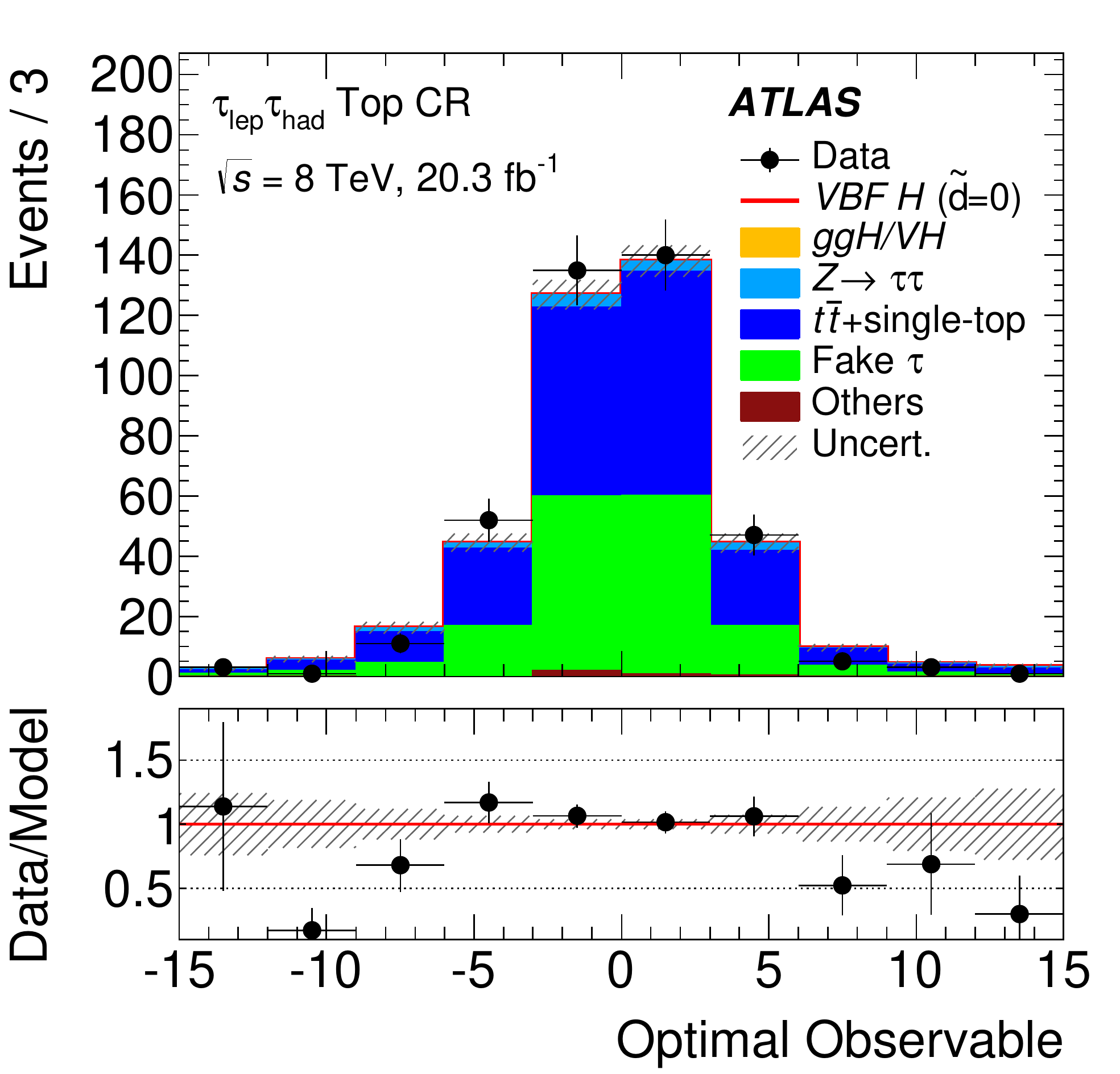} } 
   \subfloat[]{\includegraphics[width=0.49\textwidth]{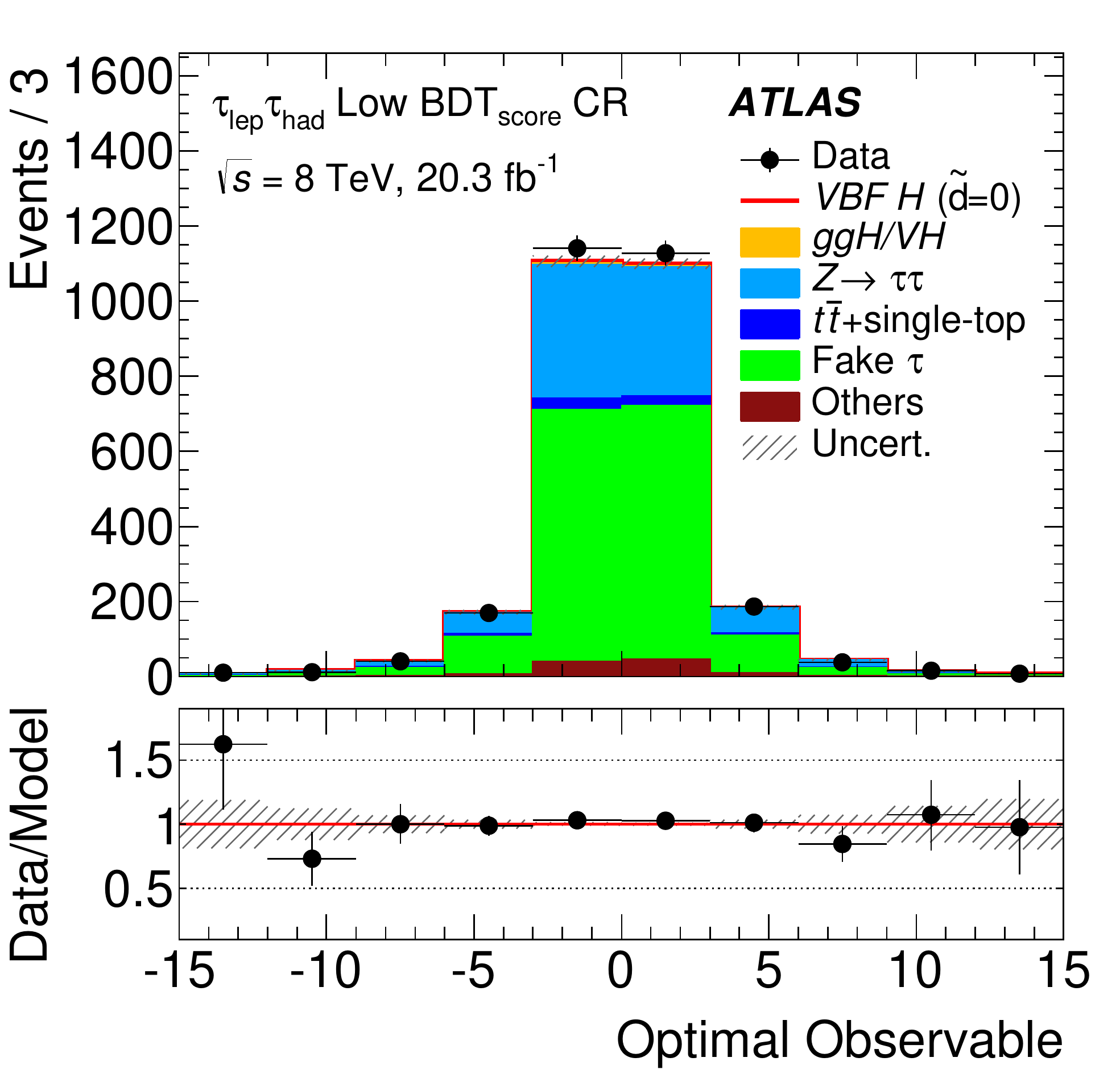} }
   \caption{Distributions of the \textit{Optimal Observable} for the \tlhad\ channel in the (a) top-quark control region (CR) and (b)  low-\bdtscore\ CR. The CR definitions are given in the text. These figures use background predictions before the global fit defined in Sect.~\ref{sec:fit}. The ``Other'' backgrounds include diboson and \Zll. Only statistical uncertainties are shown. 
} 
  \label{fig:crslh}
\end{figure}

The modelling of the \textit{Optimal Observable} distribution for various background processes is validated in dedicated control regions. 
The top-quark control regions are defined by the same cuts as the corresponding signal region, but inverting the veto on $b$-tagged jets and not applying the selection on the  \bdtscore\ (in the \tlhad\ channel a requirement of the transverse mass\footnote{The transverse mass is defined as $m_{\mathrm{T}}=\sqrt{2p_{\mathrm{T}}^{\ell} \
E_{\mathrm{T}}^{\mathrm{miss}} \cdot (1-\cos{\Delta{\phi}})}$, where
$\Delta{\phi}$ is the azimuthal separation between the directions of the
lepton and the missing transverse momentum.} $m_\mathrm{T}>40~\GeV$ is also applied). In the \tll\ channel a  \Zll\ control region is  obtained by requiring two same-flavour opposite-charge leptons, the invariant mass of the two leptons to be $80 < m_{\ell\ell} < 100~\GeV$, and no  \bdtscore\ requirement, but otherwise applying the same requirements as for the signal region. These regions are also used to normalise the respective background estimates using a global fit described in the next section.
Finally, an additional region is defined for each channel, called the low-\bdtscore\ control region, where a background-dominated region orthogonal to the signal region is selected by requiring the \bdtscore\ to be less than 0.05 for \tll\ and less than 0.3 for \tlhad.
The distribution of the \textit{Optimal Observable} in these regions is shown in Figs.~\ref{fig:crsll} and ~\ref{fig:crslh}, demonstrating the good description of the data by the background estimates. 

The effect of systematic uncertainties on the yields in signal region and on the shape of the \textit{Optimal Observable} is evaluated following the procedures and prescriptions described in Ref.~\cite{HtautauCouplings}.  
An additional theoretical uncertainty in the shape of the \textit{Optimal Observable} is included to account for the signal reweighting procedure described in Sect.~\ref{sec:samples}. This is obtained from the small difference between the \textit{Optimal Observable} distribution in reweighted samples, compared to samples with anomalous couplings directly generated with \textsc{MadGraph5\_aMC@NLO}.
While the analysis is statistically limited, the most important systematic uncertainties are found to arise from effects on the jet, hadronically decaying $\tau$ and electron energy scales; the most important theoretical uncertainty is due to the description of the underlying event and parton shower in the VBF signal sample.

\clearpage

%% file: fit.tex
\section{Fitting procedure}
\label{sec:fit}

The best estimate of \dtilde\ is obtained using a maximum-likelihood fit performed on the \textit{Optimal Observable} distribution in the signal region  for each decay channel simultaneously, with information from different control  regions included to constrain background normalisations and nuisance parameters. The normalisation of the  VBF \Htt\ and \HWW~signal sample is left free in the fit, i.e.\ this analysis only exploits the shape of the \textit{Optimal Observable} and does not depend on any possibly model-dependent information about the cross section of CP-mixing scenarios. The relative proportion of the two Higgs boson decay modes is assumed to be as in the SM. 
All other Higgs boson production modes are treated as background in this study and normalised to their SM expectation, accounting for the corresponding theoretical uncertainties.

A binned likelihood function $\mathcal{L}(\mathbf{x}; \mu, \boldsymbol{\theta})$ is employed, which is a function of the  data $\mathbf{x}$, the free-floating signal strength $\mu$, defined as the ratio of the measured cross section times branching ratio to the Standard Model prediction, 
and further nuisance parameters $\boldsymbol{\theta}$. It relies on an underlying model of signal plus background, and it is defined as the product of Poisson probability terms for each bin in the distribution of the \textit{Optimal Observable}. 
A set of signal templates corresponding to different values of the CP-mixing parameter \dtilde~is created by reweighting the SM VBF \Htt\ and \HWW~signal samples, as described in Sect.~\ref{sec:samples}. The likelihood function is then evaluated for each \dtilde~hypothesis using the corresponding signal template,  while keeping the same background model.
The calculation profiles the nuisance parameters to the best-fit values $\hat{\boldsymbol{\theta}}$, including information about systematic uncertainties and normalisation factors, both of which affect the expected numbers of signal and background events.

After constructing the negative log-likelihood (NLL) curve by calculating the NLL value for each \dtilde~hypothesis, the 
approximate central confidence interval at 68\% confidence level (CL) is determined from the best estimator $\hat{\dtilde}$, at which the NLL curve has its minimum value, by reading off the points at which $\Delta$NLL=NLL$-$NLL$_{\text{min}} = 0.5$. 
The expected sensitivity is determined 
using an Asimov dataset, i.e.\ a pseudo-data distribution equal to the signal-plus-background expectation for given values of \dtilde\ and the parameters of the fit, in particular the signal strength $\mu$, and not including statistical fluctuations~\cite{Cowan:2010js}.

In both channels, a region of low \bdtscore\ is obtained as described in the preceding section. The distribution of the \bdtscore\ itself is fitted in this region, which has a much larger number of background events than the signal region, allowing the nuisance parameters to be constrained by the data. This region provides the main constraint on the \Ztt\ normalisation, which is free to float in the fit. The event yields from the top-quark (in \tll\ and \tlhad) and \Zll\ (in \tll\ only) control regions defined in the previous section are also included in the fit, to constrain the respective background normalisations, which are also left free in the fit. 

\begin{figure}[h!]
  \centering
   \subfloat[]{\includegraphics[width=0.45\textwidth]{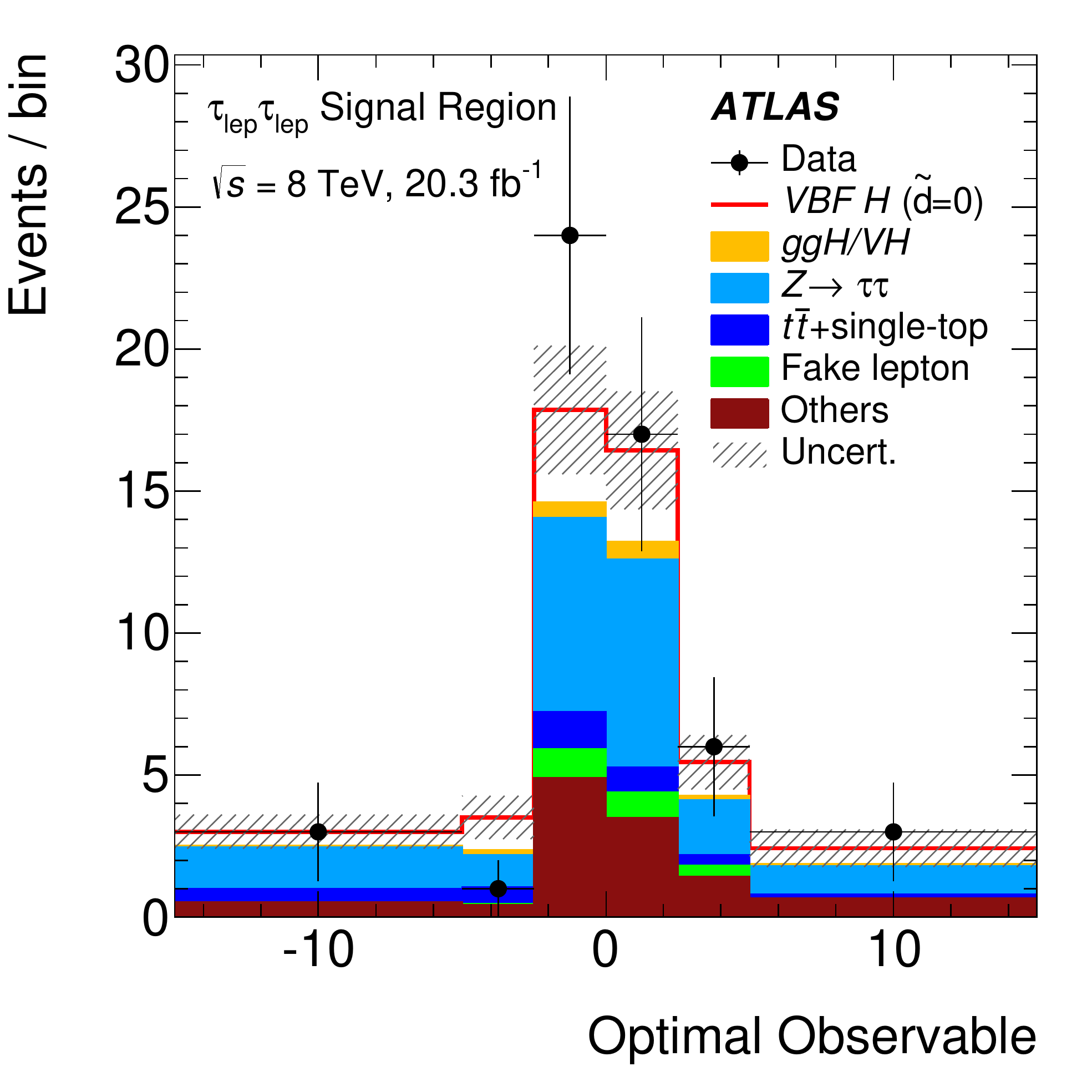}  }
   \subfloat[]{\includegraphics[width=0.45\textwidth]{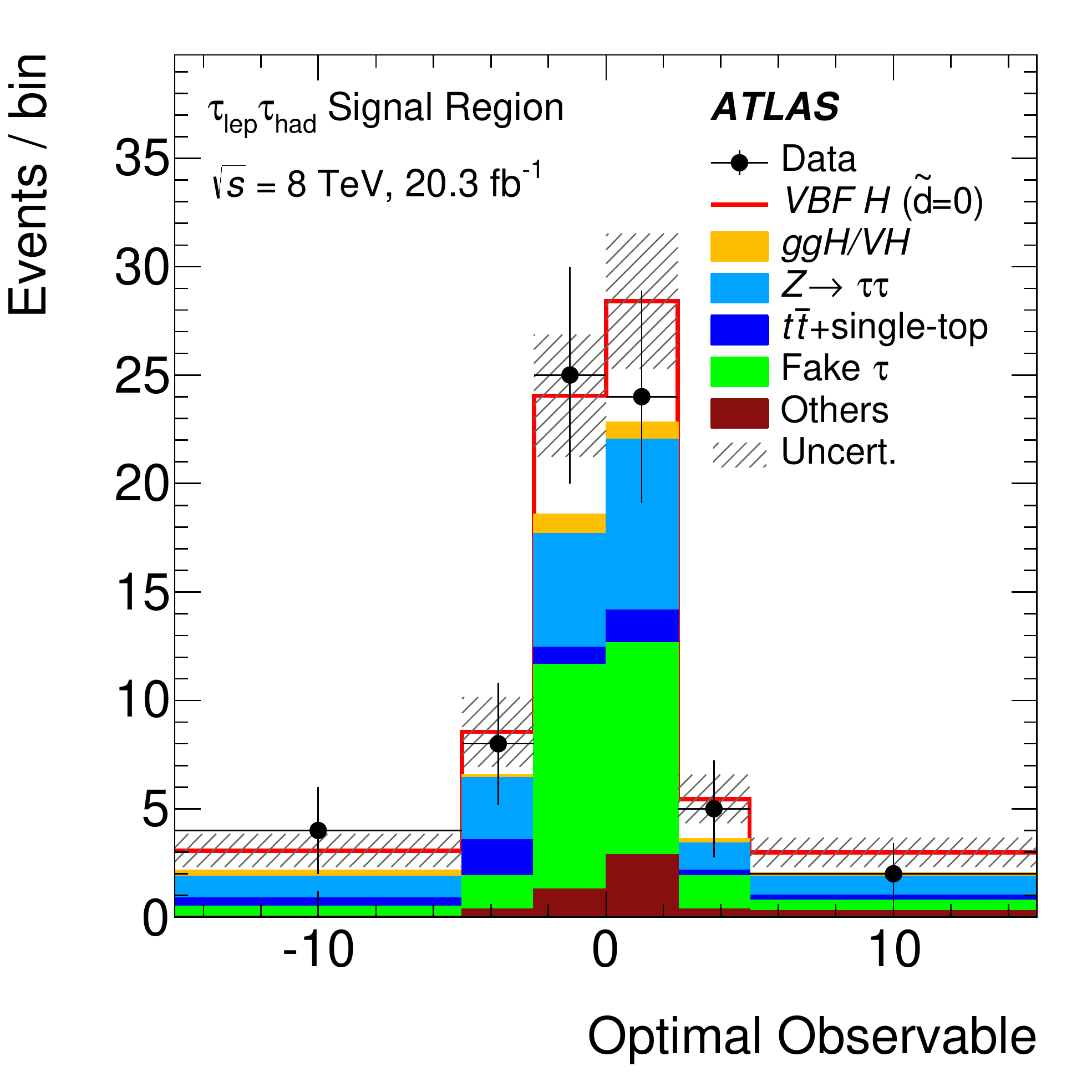} }
   \caption{Distributions of the \textit{Optimal Observable} in the signal region for the (a) \tll\  and (b) \tlhad\  channel, after the global fit performed for the $\dtilde=0$ hypothesis. The best-fit signal strength is  $\mu=1.55^{+0.87}_{-0.76}$. The ``Other'' backgrounds include diboson and \Zll. The error bands include all uncertainties.   
   }
  \label{fig:oosr}
\end{figure}

The distributions of the \textit{Optimal Observable} in each channel are shown in Fig.~\ref{fig:oosr}, with the nuisance parameters, background and signal normalisation adjusted by the global fit performed for the $\dtilde=0$ hypothesis. Table~\ref{tab:expectedyields} provides the  fitted yields of signal and background events, split into the various contributions, in each channel. The number of events  observed in data is also provided.

\begin{table}[t!]                                                                                                                           
  \begin{center}
    \begin{tabular}{|c| r@{$\pm$}l | r@{$\pm$}l |}
      \hline
      Process      & \multicolumn{2}{c|}{\tll}            &  \multicolumn{2}{c|}{\tlhad}             \\ \hline \hline
      Data         & \multicolumn{2}{c|}{54}          & \multicolumn{2}{c|}{68}              \\ \hline
            VBF \Htt/$WW$ & $9.8 $ & $ 2.1$   & $16.7 $ & $ 4.1$     \\ \hline
      \Ztt         & $19.6 $ & $ 1.0$  & $19.1 $ & $ 2.2$     \\
      Fake lepton/$\tau$     & $2.3 $ & $ 0.3 $  & $24.1 $ & $ 1.5$       \\
      \ttbar+single-top          & $3.8 $ & $ 1.0$   & $4.8 $ & $ 0.7$      \\
      Others & $11.5$ & $ 1.7$   & $5.3 $ & $ 1.6$      \\
      $ggH/VH$, $H\to\tau\tau/WW$  & $1.6 $ & $ 0.2$   & $ 2.5$ & $ 0.7 $      \\ \hline
Sum of backgrounds & $38.9 $ & $ 2.3$ & $55.8 $ & $ 3.3$   \\ \hline
    \end{tabular}
    \caption{Event yields in the signal region, after the global fit performed for the $\dtilde=0$ hypothesis. The errors include systematic uncertainties.}
    \label{tab:expectedyields}
  \end{center}                                                                                                                             
\end{table}

\clearpage

%% file: results.tex
\section{Results}
\label{sec:results}

The mean value of the \textit{Optimal Observable} for the signal is expected to be zero for a CP-even case, while there may be deviations in case of CP-violating effects. 
A mean value of zero is also expected for the background, as has been demonstrated. Hence, the mean value in data should also be consistent with zero if there are no CP-violating effects within the precision of this measurement.  
The observed values for the mean value in data inside the signal regions are $0.3\pm0.5$ for \tll\ and $-0.3\pm0.4$ for \tlhad, fully consistent with zero within statistical uncertainties 
and thus showing no hint of CP violation.

\begin{figure}[h!]
  \centering
   \includegraphics[width=0.6\textwidth]{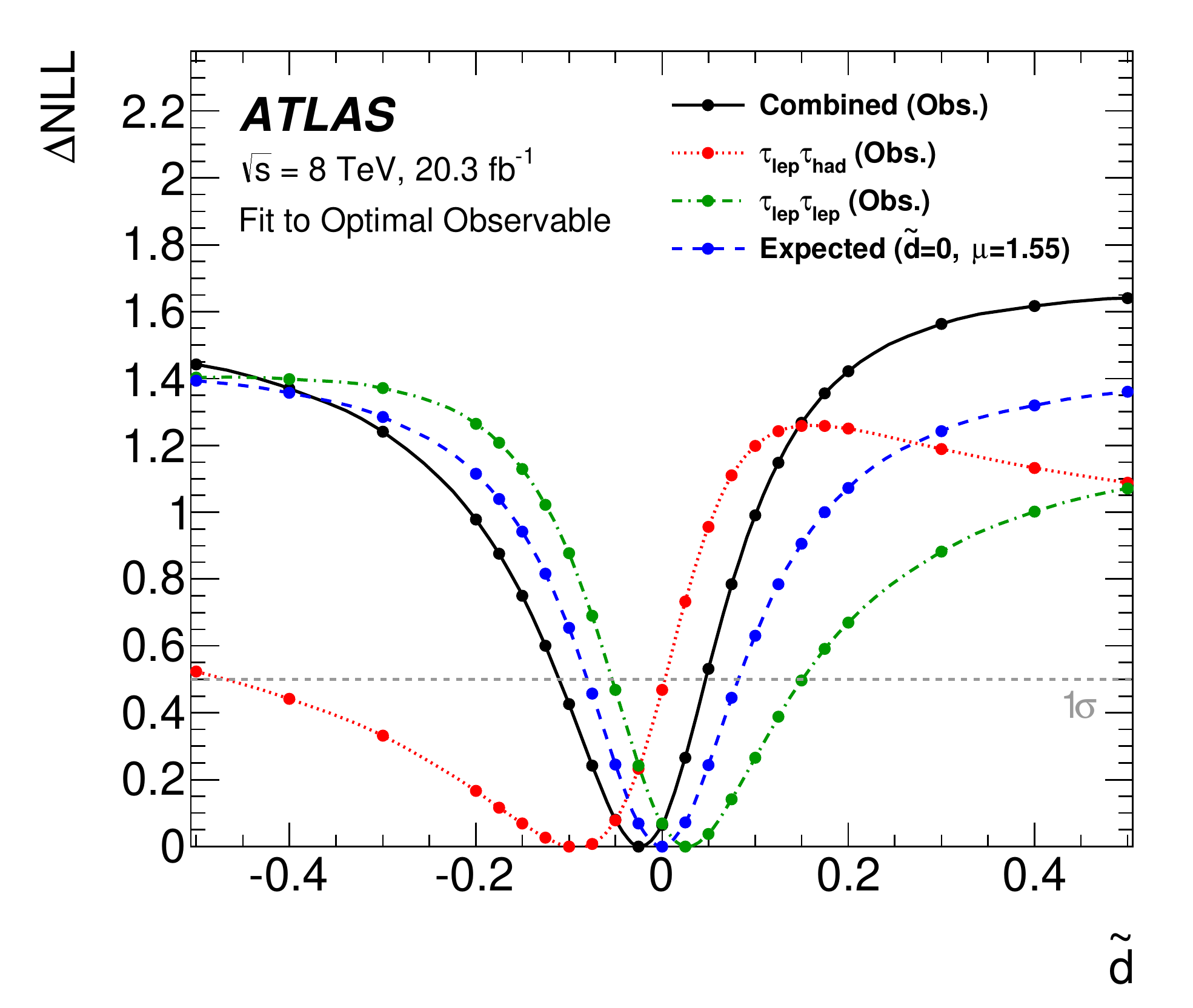}  
   \caption{Observed and expected $\Delta$NLL as a function of the \dtilde~values defining the underlying signal hypothesis, for \tll\ (green), \tlhad\ (red) and their combination (black). The best-fit values of all nuisance parameters from the combined fit at each \dtilde~point  were used in all cases. An Asimov dataset with SM backgrounds plus pure CP-even VBF signal ($\dtilde=0$), scaled to the best-fit signal-strength value, was used to calculate the expected values, shown in blue. The markers indicate the points where an evaluation was made -- the lines are only meant to guide the eye. }
  \label{fig:NLLcurve}
\end{figure}

\begin{figure}[h!]
  \centering
   \includegraphics[width=0.5\textwidth]{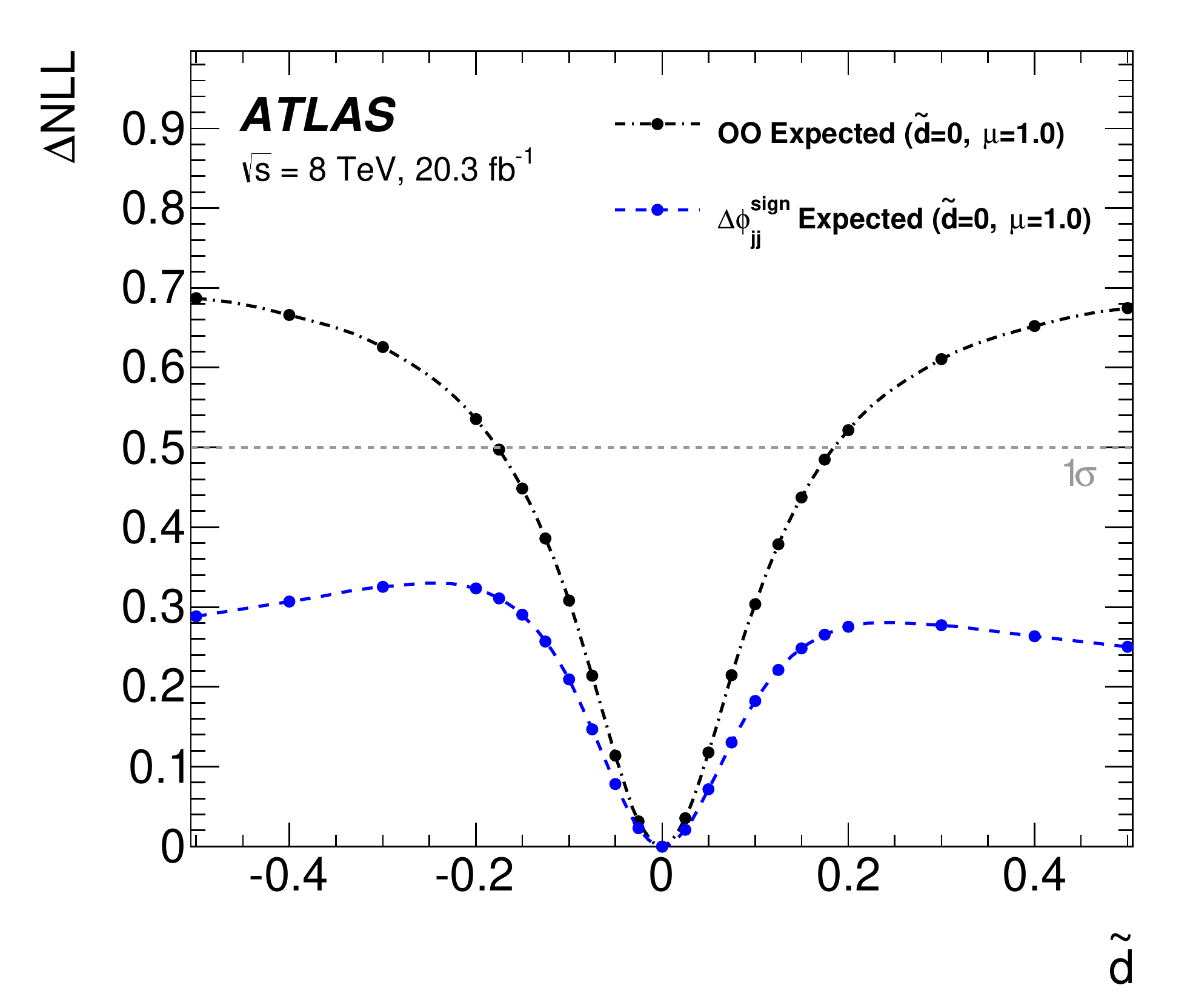}
   \caption{Expected  $\Delta$NLL for the combination of both channels as a function of the \dtilde~values defining the underlying signal hypothesis when using the  \textit{Optimal Observable} (black) or the  \signdphi\ parameter (blue) as the final discriminating variable.  An Asimov dataset with SM backgrounds plus pure CP-even VBF signal ($\dtilde=0$) scaled to the SM expectation was used to calculate the expected values in both cases. The markers indicate the points where an evaluation was made -- the lines are only meant to guide the eye.} 
  \label{fig:NLLcomparison}
\end{figure}

\begin{figure}[h!]
  \centering
   \includegraphics[width=0.5\textwidth]{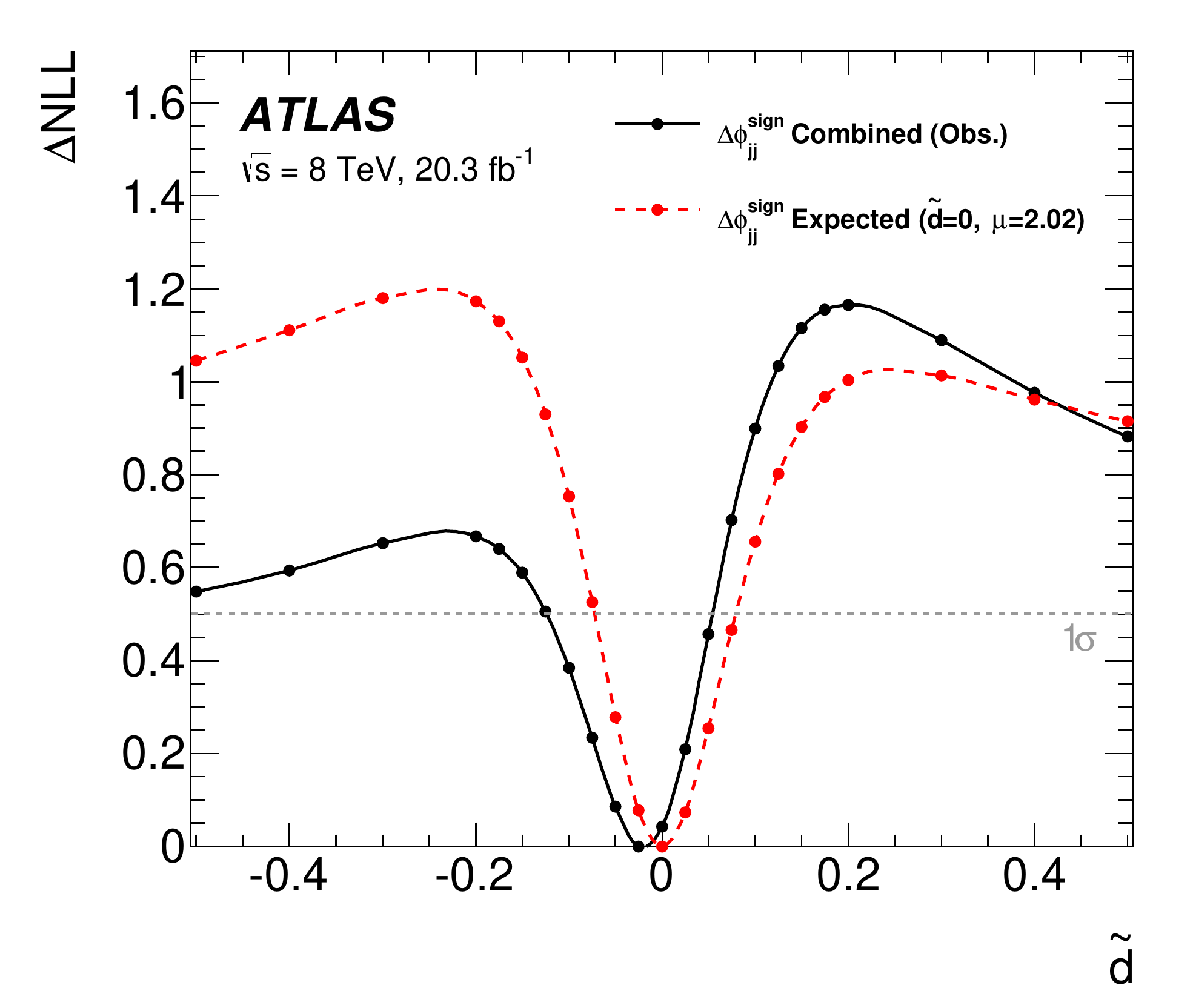}
   \caption{Observed (black) and expected (red) $\Delta$NLL for the combination of both channels as a function of the \dtilde~values defining the underlying signal hypothesis when using the  \signdphi\ parameter as the final discriminating variable. An Asimov dataset with SM backgrounds plus pure CP-even VBF signal ($\dtilde=0$), scaled to the best-fit value of the signal strength in the combined fit when using the \signdphi\ parameter ($\mu=2.02^{+0.87}_{-0.77}$) was used to calculate the expected values. The markers indicate the points where an evaluation was made -- the lines are only meant to guide the eye.} 
  \label{fig:NLLdphi}
\end{figure}

As described in the previous section, the observed limit on CP-odd couplings
is estimated using a global maximum-likelihood fit to the \textit{Optimal Observable} distributions in data.
The observed distribution of $\Delta$NLL as a function of the CP-mixing  parameter $\dtilde$ for the individual channels separately, and for
their combination, is shown in Fig.~\ref{fig:NLLcurve}. The \tll\ and \tlhad\ curves use the best-fit values of all nuisance parameters from the 
combined fit at each $\dtilde$ point. The expected curve is calculated assuming no CP-odd coupling, with the 
$\Htt$ signal scaled to the signal-strength value ($\mu = 1.55^{+0.87}_{-0.76}$) determined from the fit for $\dtilde=0$. 
In the absence of CP violation the curve is expected to have a minimum at $\dtilde=0$. Since the first-order \textit{Optimal Observable} used in the present analysis is only sensitive to small variations in the considered variable, for large $\dtilde$ values there is no further discrimination power and thus the $\Delta$NLL curve is expected to flatten out. The observed curve follows this behaviour and is consistent with no CP violation. The regions $\dtilde$ < $-0.11$ and $\dtilde$ > 0.05 are excluded at 68\% CL. The expected confidence intervals are $[-0.08 , 0.08]~([-0.18 , 0.18])$ for an assumed signal strength of 
$\mu =$ 1.55 (1.0).
The constraints on the CP-mixing parameter $\dtilde$ based on VBF production can be directly compared to those obtained by studying the Higgs boson decays into vector bosons, 
as the same relation between the $HWW$ and $HZZ$  couplings as in Ref.~\cite{Khachatryan:2014kca,Aad:2015mxa} is assumed.
The 68\% CL interval presented in this work is a factor 10 better than the one obtained in Ref. \cite{Aad:2015mxa}.

As a comparison, the same procedure for extracting the CP-mixing parameter  $\dtilde$ was applied using the \signdphi\ observable, previously proposed for this measurement and defined in Eq.~\ref{eq:signdphi}, rather than the \textit{Optimal Observable}. The expected $\Delta$NLL curves for a SM Higgs boson signal from the combination of both channels for the two CP-odd observables 
are shown in Fig.~\ref{fig:NLLcomparison}, allowing a direct comparison,  and clearly indicate the better sensitivity  of the \textit{Optimal Observable}.
The observed $\Delta$NLL curve derived from the \signdphi\ distribution  is also consistent with   $\dtilde=0$, 
as shown in Fig.~\ref{fig:NLLdphi}, along with the expectation for a signal with $\dtilde=0$ scaled to the best-fit signal-strength value ($\mu=2.02^{+0.87}_{-0.77}$).

%% file: conclusions.tex
\section{Conclusions}
\label{sec:conclusions}

A test of CP invariance in the Higgs boson coupling to vector-boson has been performed using the vector boson fusion production mode and the $H\rightarrow \tau\tau$ decay. The dataset corresponds to 20.3~\ifb of $\sqrt{s}$ = 8~\TeV\ proton--proton collisions recorded by the ATLAS detector at the LHC. Event selection, background estimation and evaluation of systematic uncertainties are all very similar to the ATLAS analysis that provided evidence of the \Htt\ decay. 
An \textit{Optimal Observable} is constructed and utilised, and is shown to provide  
a substantially better sensitivity than the variable traditionally proposed for this kind of study, \signdphi. No sign of CP violation is observed. Using only the dileptonic and semileptonic \Htt\ channels, and under the assumption $\tilde{d} = \tilde{d}_B$, values of \dtilde\ less than -0.11 and greater than 0.05 are excluded at 68\% CL. 

This 68\% CL interval is a factor of 10 better than the one previously obtained by the ATLAS experiment from Higgs boson decays into vector bosons. 
In contrast, the present analysis has no sensitivity to constrain a 95\% CL interval with the dataset currently available -- however larger data samples in the future and consideration of additional Higgs boson decay channels should make this approach highly competitive.

%% file: Acknowledgements.tex


\section*{Acknowledgements}

We thank CERN for the very successful operation of the LHC, as well as the
support staff from our institutions without whom ATLAS could not be
operated efficiently.

We acknowledge the support of ANPCyT, Argentina; YerPhI, Armenia; ARC, Australia; BMWFW and FWF, Austria; ANAS, Azerbaijan; SSTC, Belarus; CNPq and FAPESP, Brazil; NSERC, NRC and CFI, Canada; CERN; CONICYT, Chile; CAS, MOST and NSFC, China; COLCIENCIAS, Colombia; MSMT CR, MPO CR and VSC CR, Czech Republic; DNRF and DNSRC, Denmark; IN2P3-CNRS, CEA-DSM/IRFU, France; GNSF, Georgia; BMBF, HGF, and MPG, Germany; GSRT, Greece; RGC, Hong Kong SAR, China; ISF, I-CORE and Benoziyo Center, Israel; INFN, Italy; MEXT and JSPS, Japan; CNRST, Morocco; FOM and NWO, Netherlands; RCN, Norway; MNiSW and NCN, Poland; FCT, Portugal; MNE/IFA, Romania; MES of Russia and NRC KI, Russian Federation; JINR; MESTD, Serbia; MSSR, Slovakia; ARRS and MIZ\v{S}, Slovenia; DST/NRF, South Africa; MINECO, Spain; SRC and Wallenberg Foundation, Sweden; SERI, SNSF and Cantons of Bern and Geneva, Switzerland; MOST, Taiwan; TAEK, Turkey; STFC, United Kingdom; DOE and NSF, United States of America. In addition, individual groups and members have received support from BCKDF, the Canada Council, CANARIE, CRC, Compute Canada, FQRNT, and the Ontario Innovation Trust, Canada; EPLANET, ERC, FP7, Horizon 2020 and Marie Sk{\l}odowska-Curie Actions, European Union; Investissements d'Avenir Labex and Idex, ANR, R{\'e}gion Auvergne and Fondation Partager le Savoir, France; DFG and AvH Foundation, Germany; Herakleitos, Thales and Aristeia programmes co-financed by EU-ESF and the Greek NSRF; BSF, GIF and Minerva, Israel; BRF, Norway; Generalitat de Catalunya, Generalitat Valenciana, Spain; the Royal Society and Leverhulme Trust, United Kingdom.

The crucial computing support from all WLCG partners is acknowledged
gratefully, in particular from CERN and the ATLAS Tier-1 facilities at
TRIUMF (Canada), NDGF (Denmark, Norway, Sweden), CC-IN2P3 (France),
KIT/GridKA (Germany), INFN-CNAF (Italy), NL-T1 (Netherlands), PIC (Spain),
ASGC (Taiwan), RAL (UK) and BNL (USA) and in the Tier-2 facilities
worldwide.

%% file: atlas_authlist.tex
\begin{flushleft}
{\Large The ATLAS Collaboration}

\bigskip

G.~Aad$^\textrm{\scriptsize 87}$,
B.~Abbott$^\textrm{\scriptsize 114}$,
J.~Abdallah$^\textrm{\scriptsize 152}$,
O.~Abdinov$^\textrm{\scriptsize 11}$,
B.~Abeloos$^\textrm{\scriptsize 118}$,
R.~Aben$^\textrm{\scriptsize 108}$,
M.~Abolins$^\textrm{\scriptsize 92}$,
O.S.~AbouZeid$^\textrm{\scriptsize 138}$,
N.L.~Abraham$^\textrm{\scriptsize 150}$,
H.~Abramowicz$^\textrm{\scriptsize 154}$
% ...
\bigskip
\\
% ...
$^{11}$ Institute of Physics, Azerbaijan Academy of Sciences, Baku, Azerbaijan\\
% ...
$^{87}$ CPPM, Aix-Marseille Universit{\'e} and CNRS/IN2P3, Marseille, France\\
% ...
$^{92}$ Department of Physics and Astronomy, Michigan State University, East Lansing MI, United States of America\\
% ...
$^{108}$ Nikhef National Institute for Subatomic Physics and University of Amsterdam, Amsterdam, Netherlands\\
% ...
$^{114}$ Homer L. Dodge Department of Physics and Astronomy, University of Oklahoma, Norman OK, United States of America\\
% ...
$^{118}$ LAL, Univ. Paris-Sud, CNRS/IN2P3, Universit{\'e} Paris-Saclay, Orsay, France\\
% ...
$^{138}$ Santa Cruz Institute for Particle Physics, University of California Santa Cruz, Santa Cruz CA, United States of America\\
% ...
$^{150}$ Department of Physics and Astronomy, University of Sussex, Brighton, United Kingdom\\
% ...
$^{152}$ Institute of Physics, Academia Sinica, Taipei, Taiwan\\
% ...
$^{154}$ Raymond and Beverly Sackler School of Physics and Astronomy, Tel Aviv University, Tel Aviv, Israel\\
% ...
\end{flushleft}
